\documentclass[twocolumn]{aastex631}
\begin{document}

\newcommand{\msun}{{M_\odot}}
\newcommand{\tess}{{TESS}}
\newcommand{\gaia}{{\em Gaia}}
\newcommand{\rapid}{{{\tt rapid}}}
\newcommand{\delayed}{{{\tt delayed}}}
\newcommand{\porb}{{{P_{\rm{orb}}}}}
\newcommand{\metal}{{{\rm Z}}}

\newcommand{\mlc}{{{M_{\rm{LC}}}}}
\newcommand{\llc}{{{L_{\rm{LC}}}}}
\newcommand{\myr}{{{\rm Myr}}}
\newcommand{\mbh}{{{M_{\rm{BH}}}}}
\newcommand{\mco}{{{M_{\rm{CO}}}}}
\newcommand{\mtot}{{M_{\rm{tot}}}}
\newcommand{\ecc}{{{Ecc}}}
\newcommand{\yr}{{{\rm{yr}}}}
\newcommand{\kpc}{{{\rm{kpc}}}}
\newcommand{\pc}{{{\rm{pc}}}}
\newcommand{\BH}{{\rm{BH}}}
\newcommand{\LC}{{\rm{LC}}}
\newcommand{\cosmic}{{{\texttt{COSMIC}}}}
\newcommand{\mwdust}{{\texttt{mwdust}}}
\newcommand{\bse}{{{\texttt{BSE}}}}
\newcommand{\Mmin}{{M_{\rm{min}}}}
\newcommand{\Mmax}{{M_{\rm{max}}}}
\newcommand{\sma}{{a}}
\newcommand{\fallback}{{\rm{FB}}}
\newcommand{\bhlc}{{\rm{BH\text{--}LC}}}
\newcommand{\mtwelve}{{\bf{m12i}}}
\newcommand{\beam}{{{\rm{RB}}}}
\newcommand{\kms}{{\rm{km\ s^{-1}}}}
\newcommand{\tdur}{{t_{\rm{dur}}}}
\newcommand{\snr}{{{\rm{SNR}}}}
\newcommand{\days}{{\rm{day}}}
\newcommand{\qmin}{{q_\mathrm{min}}}

\title{{\bf Detecting Detached Black Hole binaries through Photometric Variability}}
\author[0000-0001-9685-3777]{Chirag Chawla}
\affiliation{Tata Institute of Fundamental Research, Department of Astronomy and Astrophysics, Homi Bhabha Road, Navy Nagar, Colaba, Mumbai, 400005, India}
\email{chirag.chawla@tifr.res.in}

\author[0000-0002-3680-2684]{Sourav Chatterjee}
\affiliation{Tata Institute of Fundamental Research, Department of Astronomy and Astrophysics, Homi Bhabha Road, Navy Nagar, Colaba, Mumbai, 400005, India}
\email{souravchatterjee.tifr@gmail.com}

\author[0000-0002-8465-8090]{Neev Shah}
\affiliation{Indian Institute of Science Education and Research Pune, Dr. Homi Bhabha Road,
Pune 411008, India}

\author[0000-0001-5228-6598]{Katelyn Breivik}
\affiliation{Department of Physics, McWilliams Center for Cosmology and Astrophysics, Carnegie Mellon University, 5000 Forbes Avenue, Pittsburgh, PA 15213, USA}

\begin{abstract}

Understanding the connection between the properties of black holes (BHs) and their progenitors is interesting in many branches of astrophysics. 
Discovering BHs in detached orbits with luminous companions (LCs) promises to help create this map since the LC and BH progenitor are expected to have the same metallicity and formation time. 
We explore the possibility of detecting BH--LC binaries in detached orbits using photometric variations of the LC flux, induced by tidal ellipsoidal variation, relativistic beaming, and self-lensing. We create realistic present-day populations of detached BH--LC binaries in the Milky Way (MW) using binary population synthesis where we adopt observationally motivated initial stellar and binary properties, star formation history and present-day distribution of these sources in the MW based on detailed cosmological simulations. We test detectability of these sources via photometric variability by \gaia\ and \tess\ missions by incorporating their respective detailed detection biases as well as interstellar extinction. We find that \gaia\ is expected to resolve  $\sim300$--$1,000$ ($\sim700$--$1,500$) detached BH--LC binaries with SNR $\ge10\ (1)$ depending on the photometric precision and details of supernova physics. Similarly, the number of resolved BH--LC binaries with TESS are $\sim50$--$200$ ($\sim140$--$350$). We find that $136^{+15}_{-15}$ BH--LC binaries would be common between \gaia\ and TESS. Moreover, between $\sim60$--$70$ ($\sim50$--$200$) BH--LC binaries identifiable using photometry with SNR $\ge10$ may also be resolved using \gaia's radial velocity (astrometry).

\end{abstract}

\section{Introduction}
\label{s:intro}
Recent observations of merging double compact object (CO) binaries by the LIGO-Virgo-KAGRA (LVK) detectors have reignited the interest to understand the astrophysical origins of CO binaries \citep[][]{Abbott2016a, Abbott2016b,GWTC1,GWTC2,GWTC2-pop}. A variety of ongoing and upcoming missions including the Zwicky Transient Facility \citep[ZTF,][]{Bellm_2018}, the Rubin Observatory's Legacy Survey of Space and Time \citep[LSST,][]{LSST}, the All-Sky Automated Survey for Supernova \citep[ASAS-SN,][]{Shappee_2014,Kochanek_2017}, SDSS-V \citep[][]{SDSS-V} and eROSITA \citep{eROSITA} are expected to unravel hundreds to thousands of COs including Cataclysmic variables, Supernova explosions, Gamma-ray bursts, and X-ray binaries. Although understanding the demographics of dark remnants in general, and BHs in particular, is interesting for many branches of astrophysics, a model independent map connecting BHs to their stellar progenitors remains elusive due to challenges in the detailed theoretical modeling of the supernova physics \citep[][]{Patton2020,Patton_2022,Fryer2022} and scarcity of discovered BHs where constraints to the progenitor properties are available \citep[e.g.,][]{Breivik2019,El-badry_2022e}. 

BH--LC binaries in detached orbits, discovered in large numbers, can be instrumental in improving this gap in our understanding of the details of how high-mass stars evolve, explode, and form COs \citep[][]{Breivik_2017,Chawla2021,Shikauchi_2023}. In particular, if the distance to the LC is known (e.g., via \gaia\ astrometry), meaningful constraints can be placed on the metallicity and age of the LC (and thus the BH's progenitor) through stellar evolution models, asteroseismology, and spectroscopy \citep[e.g.][]{Lin_2018,Angus_2019,Bellinger_2019}. 
It is expected that $\sim10^{8}-10^{9}$ stellar-mass BHs are present in the present-day MW \citep[][]{Brown1994,Olejak2020,Sweeney2022}. 
Of these, roughly $10^{4}-10^{5}$ are expected to be in binaries with a non-BH. An overwhelming $70-98\%$ of BH--LC binaries are expected to have the LC in a detached orbit. In contrast, BHs in potentially mass transferring systems is only $2-30\%$ \citep[][]{Breivik_2017,Wiktorowicz2019,Chawla2021}.
Recent advances in time-domain astronomy promises to provide unprecedented constraints on BHs in detached orbits via high-precision astrometric, photometric, and spectroscopic measurements. BHs can be characterized using several techniques: (a) astrometrically constraining the orbit of a LC by observing its motion around an unseen primary \citep[][]{Kamp_1975,Gould2002,Tomsick_2010}, (b) spectroscopically measuring the RV of the LC as it moves around the BH \citep[][]{Zeldovich_1966,Trimble_1969}, and (c) phase-curve analysis of orbital photometric modulations of the LC induced by its dark companion \citep[][]{Shakura_1987,KHRUZINA_1988}. Of course, at least for some sources, a combination of more than one of the above methods may become useful. 

The prospect of detecting BHs in detached BH--LC binaries via ongoing astrometry and RV surveys like \gaia\ and LAMOST has been extensively explored \citep[][]{Barstow2014,Breivik_2017,Mashian2017,Chawla2021,Janssens2022}. Although the estimated number of detectable BH--LC binaries is model dependent (because of model uncertainties, e.g., in SNe physics), all of these studies predict that \gaia\ could possibly discover $10-10^{3}$ BH--LC binaries in detached orbits during its $10$ year mission. In addition, the knowledge of the stellar parameters like luminosity, age, and mass of the LC can help constrain the mass of the BH as well as its progenitor's properties in a model independent way \citep[][]{Fuchs2005,Andrews_2019,Shahaf_2019,Chawla2021}.

Several non-interacting BH--LC candidates have been discovered in star clusters \citep[][]{Giesers2018,Giesers2019} and in the Large Magellanic Cloud \citep{Shenar2022a, Lennon2021,Saracino2021,Shenar2022b}. In the Galactic field also, several discoveries of candidate BH--LC binaries in detached orbits, including BHs in triples, have been proposed by studies using photometric and spectroscopic observations \citep[][]{Qian_2008,Casares_2014,Khokhlov_2018,Liu2019,Thompson2019,Rivinius_2020,Gomez_2021,Jayasinghe2021}. However, significant debate persists on the candidature of many of these systems \citep[e.g., it has been suggested that the unseen companion may actually be a low-luminosity sub-giant companion or stellar binary instead of a BH in some of the candidate systems;][]{van-den-Heuvel2020,El-Badry2021,El-Badry2022a,El-Badry2022b,El-Badry2022c}. 

Most recently, using \gaia's DR3 several groups have reported a number of possible dormant CO--LC candidate binaries using astrometry \citep[][]{Andrews_2022a,Arenou_2023,Chakrabarti_2023,El-badry_2022e,El-badry_2023,Shahaf_2022}, photometry \citep[][]{Gomel_2023} and spectroscopy \citep[][]{Fu_2022,Jayasinghe_2023,Nagarajan_2023,Tanikawa_2023,Zhao_2023}, which indicates that indeed a large population of BH--LC binaries in detached orbits do exist in nature and are waiting to be found.  

Observations of photometric variability in stars due to planetary transits have revolutionized the field by detecting thousands of exoplanets from various wide-field ground-based missions including HAT \citep[][]{Bakos_2004}, TrES \citep[][]{Alonso_2004}, XO \citep[][]{McCullough_2005}, WASP \citep[][]{Pollacco_2006} and KELT \citep[][]{Pepper_2007} and space missions like CoRoT \citep[][]{Auvergne_2009}, {\em{Kepler}} \citep[][]{Borucki_2011}, and TESS \citep[][]{Ricker2014}. Periodic variability in the observed LC flux is also expected in compact orbits around BHs. For example, in a compact enough orbit to a BH, the sky-projected surface area of an LC may show orbital phase-dependent changes resulting in the so-called ellipsoidal variations (EV) in the total observed flux. In addition, relativistic beaming (RB) of the LC's light may be strong enough to be detectable if its orbit is close-enough to a BH. Furthermore, if the geometry is favorable, light from the LC may be lensed by the BH, and the magnification due to this self-lensing (SL) within the binaries may be large enough to be detectable. Stellar binaries have already been detected using such photometric variations both in eclipsing and non-eclipsing configurations \citep[][]{Morris_1985,Thompson_2012,Herrero_2014,Nie_2017}. While microlensing surveys such as OGLE and MACHO have reported a number of isolated compact object candidates \citep[][]{Abdurrahman_2021,Lam_2022,Mroz_2022,Sahu_2022}, detection of compact objects in orbit around an LC remains illusive. Nevertheless, recent theoretical studies estimate that a significant number of detached BH--LC binaries ($\sim10-100$) may be observed by phase-curve analysis of their LCs with ongoing photometric surveys such as \tess, ZTF, LSST and {\em Kepler} \citep[][]{Masuda2019,Gomel_2021a,Wiktorowicz2021,Hu_2023}.

Using our realistic simulated Galactic populations of BH--LC binaries presented in the context of \gaia's astrometric detectability \citep[][hereafter Paper I]{Chawla2021}, we investigate the ability of \gaia\ and TESS to resolve detached BH--LCs via photometric orbital modulation induced through tidal and relativistic effects. A similar analysis for LSST would also be very interesting, however, a realistic analysis for signal to noise ratio (SNR) is not straightforward for LSST at this time. We discuss the details of our simulated models in \autoref{S:methods}. In \autoref{S:transit-snr} we describe how we calculate SNR taking into account the detection biases and our adopted detection criteria. In \autoref{S:results} we present our key results for the intrinsic as well as detectable BH--LCs. We discuss possibilities from follow-up studies in \autoref{S:follow-up} and conclude in \autoref{S:conclusions}. 

\section{Numerical methods}
\label{S:methods}
The synthetic populations used in this study are described in detail in Paper\ I. Nevertheless, we present the crucial details relevant for this study for completeness. We create representative present-day BH--LC populations using the state-of-art Python-based rapid binary population synthesis (BPS) suite \cosmic\ \citep[][]{Breivik2020} which employs the \texttt{SSE}/\bse\ evolutionary framework to evolve single and binary stars \citep[][]{Hurley2000,Hurley2002}. 

Using \cosmic\ we generate a population of zero-age main sequence (ZAMS) binaries by assigning initial ages, metallicities ($\metal$), masses, semimajor axes ($\sma$), and eccentricities ($\ecc$). The initial age and metallicity of each binary is sampled from the final snapshot of the \textbf{m12i} model galaxy from the Latte suite of the Feedback In Realistic Environments (FIRE-2) simulations \citep[][]{Wetzel2016,Hopkins2018}. The single-star stellar evolution tracks used in \cosmic\ only incorporate metallicities in the range $\rm{\log(\metal/\metal_{\odot})}\ =\ -2.3$ and $0.2$, where $\metal_{\odot}=0.02$ is the solar metallicity. Hence, we confine the metallicities of our simulated binaries within this range and assign the limiting values for metallicities in \textbf{m12i} which are outside this range.

The stellar and orbital parameters for the ZAMS population such as mass, orbital period ($\porb$), $\ecc$ and mass ratio ($q\leq1$) are sampled from observationally motivated probability distribution functions. The primary mass is sampled from the  \citet[][]{Kroupa2001} initial stellar mass function (IMF) between $\Mmin/\msun=0.08$ and $\Mmax/\msun=150$ and the secondary mass is assigned in the range $\Mmin$ to the primary mass using a flat $q$ distribution \citep[][]{Mazeh1992,Goldberg1994}. We assume initially thermal $\ecc$ distribution \citep[e.g.,][]{Jeans_1919,Ambartsumian_1937,Heggie1975}. The initial $\sma$ are drawn to be uniform in log with an upper bound of $10^{5}{R_{\odot}}$ and inner bound such that $R_{\rm{pri}}\le R_{\rm{RL}}/2$ \citep[][]{Han1998}, where $R_{\rm{pri}}$ is the radius of the primary component and $R_{\rm{RL}}$ is its Roche radius.

\cosmic\ uses several modified prescriptions beyond the standard \bse\ implementations described in \citet{Hurley2002} to evolve the binary population from ZAMS to get the present-day BH--LC population in the MW. For a detailed description of these modifications see \citet{Breivik2020}. The properties of the present day BH--LC binary population depends strongly on the outcome of the Roche-overflow mass transfer from the BH progenitor and the natal kick imparted during BH formation. 
We adopt critical mass ratios as a function of donor type derived from the adiabatic response of the donor radius and its Roche radius \citep[][]{Belczynski2008} to determine whether mass transfer remains dynamically stable or leads to a common envelope (CE) evolution \citep{Webbink_1985}. 

For CE evolution \cosmic\ uses a formulation based on the orbital energy; the CE is parameterized using two parameters $\alpha$ \citep[][]{Livio_1984} and $\lambda$ where, $\alpha$ denotes the efficiency of using the orbital energy to eject the envelope and $\lambda$ defines the binding energy of the envelope based on the donor's stellar structure \citep{Tout1997}. We adopt $\alpha=1$ and that $\lambda$ depends on the evolutionary phase of the donor \citep[see Appendix A of][]{Claeys2014}. We adopt two widely used explosion mechanisms for BH formation via core-collapse supernovae (CCSNe): ``rapid" and ``delayed" \citep[][]{Fryer2012}. While these two prescriptions introduce several differences in the BHs' birth mass function and natal kicks, the most prominent for our study is the presence (absence) of a mass gap between ($3$ and $5\,M_\odot$) NSs and BHs produced via core-collapse SNe (CCSNe) in the rapid (delayed) prescription. We refer to the model populations created using the rapid (delayed) prescription as \rapid\ (\delayed) model. We assign BH natal kicks with magnitude $v(1-f_\fallback)$, where $v$ is drawn from a Maxwellian distribution with $\sigma=265\,$ \citep[][]{Hobbs2005} and $f_\fallback$ is the fraction of mass fallback from the outer envelope of the exploding star \citep[e.g.][]{Belczynski2008}. 

\subsection{Synthetic Milky-Way population}
\label{S:synthetic-mw}
Using \cosmic, we evolve binaries until the distribution of present-day properties for BH--LC binaries saturate to a high degree of accuracy (for a more detailed description see Paper I). Typically, while creating this representative population we only simulate the evolution of a fraction of the total MW mass as defined by galaxy \textbf{m12i}. In order to obtain the correct number of BH--LC binaries in the MW at present, we up-scale the simulated BH--LC population by a factor proportional to the ratio of the entire stellar mass of {\bf{m12i}} ($M_{\bf{m12i}}$) and the total simulated single and binary stellar mass ($M_{\rm{sim}}$) in \cosmic.\footnote{We do not simulate single stars. Instead we adopt an observationally motivated binary fraction \citep[][]{Duchene_2013,Moe_2017} to estimate the equivalent total mass from the simulated binary mass.} The total number of the present-day BH--LC binaries in the MW is then defined as
\begin{equation}
   N_{\bhlc,\rm{MW}} = N_{\bhlc,\rm{sim}}\frac{M_{\rm{\bf m12i}}}{M_{\rm{sim}}}.
\end{equation}
To produce a MW-representative population of present-day BH--LC binaries, we sample (with replacement) $N_{\bhlc,\rm{MW}}$ binaries from our simulated population of BH--LC binaries to assign a complete set of stellar and orbital parameters including mass, metallicity, luminosity, radius, eccentricity, and $\porb$. Each binary is also assigned a Galactocentric position by locating the star-particle closest to the given binary in age and metallicity in the \mtwelve\ galaxy model with an offset following the Ananke framework \citep[][]{Sanderson2020}. Hence, we preserve the complex correlations between Galactic location, metallicity, and stellar density in the MW. Finally, each binary is assigned a random orientation by specifying the Campbell elements, inclination ($i$) with respect to the line of sight, argument of periapsis ($\omega$) and longitude of ascending node ($\Omega$). We create 200 MW realizations for each model (\rapid\ and \delayed) to investigate the variance associated with the random assignments of each binary in the procedure described above.

We incorporate the effect of interstellar extinction and reddening by calculating extinction ($A_v$) using the python package \mwdust\ \citep[][]{Drimmel2003,Marshall2006,Bovy2016,Green2019} based on the position of binaries in the galaxy and include this correction in estimating the TESS and \gaia\ magnitudes.

\section{Photometric Variability and Detection}
\label{S:transit-snr}

\tess\ and \gaia\ are both all-sky surveys despite different primary observing goals and strategies. For both, the number and epochs of observations of any particular source over the full mission duration are dependent on its Galactic coordinates. 
We consider three physical processes which can introduce orbital modulation in the LC's observed flux: 
ellipsoidal variation due to tidal distortion of the LC (EV), relativistic beaming (RB), and self-lensing (SL). Below we describe how we estimate the signal to noise ratio in TESS and \gaia\ photometry and our detectability conditions.

\subsection{SNR calculation}

The relevant SNR for TESS observation for a source undergoing photometric variations can be written as \citep[][]{Sullivan2015}
\begin{eqnarray}
\label{Eq:SNR_tess}
    \left[S/N\right]_{TESS} = \frac{\sqrt{\frac{1}{2\pi}\int \left(\frac{\Delta \mathcal{F}}{\langle\mathcal{F}\rangle}\right)^{2} d \phi}}{\sigma_{30}/\sqrt{N}}.
\end{eqnarray}
$\Delta \mathcal{F}\equiv\mathcal{F}(\phi) - \langle\mathcal{F}\rangle$, where $\langle\mathcal{F}\rangle$ 
is the orbital phase $\phi$-averaged flux, $\sigma_{30}$ is the per-point combined differential 
photometric precision of TESS with $30$-minute cadence. $\sigma_{30}$ depends on the source's reddening corrected TESS magnitude and calculated using the python package \texttt{ticgen} \citep{ticgen_2017, Stassun_2018}. $N$ is the number of photometric data points with $30$-minute cadence for the source given its Galactic location using the tool, \texttt{ tess-point} \citep[][]{tess_point_2020}. Similarly, for \gaia\ we define the SNR as
\begin{eqnarray}
\label{Eq:SNR_gaia}
    \left[S/N\right]_{Gaia} = \frac{\sqrt{\frac{1}{2\pi}\int \left({\Delta G(\phi)}\right)^{2} d \phi}}{\sigma_{G}/\sqrt{N}},
\end{eqnarray}

where $\Delta G \equiv G(\phi) - \langle G \rangle$, where $\langle G \rangle$ is $\phi$-averaged \gaia\ magnitude. $G(\phi)$ is calculated from $\mathcal{F}(\phi)$ and effective temperature ($T$). $\sigma_{G}$ is the photometric precision of \gaia\ with $8-10$ sec cadence and depends on $G$.\footnote{The slight difference between the SNR expressions stems from the differences in the dimensions of reported $\sigma_{30}$ and $\sigma_G$. }
$N$ is the location-dependent number of data points obtained during \gaia's $10$-year observation estimated for each source using \gaia's Observation Forecast Tool (GOST)\footnote{https://gaia.esac.esa.int/gost/}.  The magnitudes for the photometric variability of course depend on the physical process responsible. Note that the noise modelling is based on the power-law fitting. Other factors such as stellar crowding could also contribute to the SNR and hence detections. However, including a crowding model in this study would require the information of field stars from other observational catalogs which is far from the scope of this study. Also while estimating the SNR of the SL signal, an alternative approach could be to do a summation on the integrals (\autoref{Eq:SNR_tess},\ref{Eq:SNR_gaia}) on representative cadences for the mission in consideration.
 
\subsubsection{Ellipsoidal Variation}
\label{S:SNR-EV}
The tidal force of the BH distorts the shape of the LC, elongating it along the line joining the center of the BH--LC system. The surface flux distribution also changes due to gravity darkening \citep[][]{Zeipel_1924, Kopal_1959}. Due to the orbital motion of the LC, its net sky-projected area varies resulting in a periodic modulation of the observed flux. The observed flux as a function of phase ($\phi$) due to EV is \citep[][]{Morris_1993}-
\begin{eqnarray}
\label{eq:EV-lumin}
    \mathcal{F(\phi)} & = & \mathcal{F}_{0}[1+ \left(\frac{\alpha}{9} \right) \left(\frac{R_{\rm{LC}}}{a}\right)^{3}(2+2q)(2-3\sin^{2}i)\nonumber\\ 
    & + & \left(\frac{\alpha}{9}\right)\frac{1+e\ \cos\phi}{1-e^{2}} \left(\frac{R_{\rm{LC}}}{a}\right)^{3}(3q)(2-3\sin^{2}i)\\ 
    & - & (\alpha)\frac{1+e\ \cos\phi}{1-e^{2}}\left(\frac{R_{\rm{LC}}}{a}\right)^{3}(q)(\sin^{2}i)(\cos(2\omega+2\phi-\pi))]\nonumber,
\end{eqnarray}
where, $\mathcal{F(\phi)}$ represents the Flux of the LC as a function of the orbital phase ($\phi$), $\mathcal{F}_{0}$ denotes the luminosity in the absence of the BH, and $\alpha$ is  defined as 
\begin{equation}
    \alpha = \frac{15u(2+\tau)}{32(3-u)},
\end{equation}
where $u$ and $\tau$ are the limb and gravity darkening coefficients, respectively \citep[][]{Morris_1993, Engel_2020}. For simplicity, we ignore the contribution from limb and gravity darkening in this study (i.e., $u=0.3$, $\tau=0.4$, $\alpha=0.125$) since they are expected to have a small effect on the overall results. 

\subsubsection{Relativistic Beaming}
\label{S:SNR-beaming}

The photometric modulation due to the relative motion between the LC and the observer is known as relativistic beaming. The radial component of the orbital motion of the LC induced by the BH causes a phase-dependent flux variation due to relativistic effects, such as the doppler effect, time dilation, and aberration of light \citep[][]{van_Kerkwijk_2010,Bloemen_2011}. The amplitude of the photometric variation resulting from beaming is proportional to the radial velocity semi-amplitude of the LC ($K_{\rm{LC}}$), and can be expressed as \citep[][]{Loeb_2003} 
\begin{eqnarray}
\label{eq:RB-lumin}
    \frac{\Delta \mathcal{F}}{\langle\mathcal{F}\rangle} & = & 4 \alpha_{\rm{RB}} \frac{K_{LC}}{c} \nonumber\\
    & = & 2830\ \alpha_{\rm{RB}}\sin i
    \left(\frac{\porb}{\rm{days}}\right)^{-1/3}\\
    & \times & \left(\frac{\mbh+\mlc}{M_{\odot}}\right)^{-2/3}\left(\frac{\mbh}{M_{\odot}}\right), \nonumber
\end{eqnarray}
where, $\alpha_{\beam}$ is obtained by integrating the frequency-dependent term 
\begin{equation}
    \alpha_{\rm{RB},\nu} = \frac{1}{4}\left(3-\frac{d\ \ln\mathcal{F}_{\nu}}{d\ \ln\nu}\right) 
\end{equation}
over the frequency range of the band-pass \citep[][]{Loeb_2003}. The value of bolometric $\alpha_{\rm{RB}}=1$ and it deviates from $1$ while considering only a bandpass of wavelength range \citep[][]{Loeb_2003,Zucker_2007}. In the black-body approximation for a wide range of surface temperatures the value of $\alpha_{\rm{RB}}$ remains close to 1 \citep[][]{Shporer_2017}. In this study, we have assumed $\alpha_{\beam}\approx1$ for simplicity. 

\subsubsection{Self Lensing}
\label{S:SNR-self-lensing}

For BH--LC binaries with orbits aligned with the line-of-sight, the BH acts as a lens magnifying the LC thus producing a sudden shift in its luminosity during occultation. This generates a periodic spike or self-lensing signal every time the BH eclipses its companion \citep[][]{Leibovitz1971,Maeder1973,Gould_1995,Rahavar_2010}. The amplitude of the modulation in the light-curve depends on the magnification factor which is a function of binary separation, BH mass, and the inclination of the binary orbit with respect to the sky plane. The amplitude of the SL signal is adopted from \citet[][]{Witt_1994} as
\begin{equation}
     \mu_{SL} = \frac{1}{\pi}[c_{F}F(k)+c_{E}E(k)+c_{\Pi}\Pi(n,k)]
\end{equation}
where $F$, $E$, and $\Pi$ are complete elliptic integrals of first, second and third kind and the coefficients $c_{F}$, $c_{E}$, and $c_{\Pi}$ are defined as
\begin{eqnarray}
    c_{F}\ & =\ & -\frac{b-r}{r^{2}} \frac{4+(b^2-r^2)/2}{\sqrt{4+(b-r)^2}}\nonumber\\
    c_{E}\ & =\ & \frac{b+r}{2r^2}\sqrt{4+(b-r)^2}\nonumber\\
    c_{\Pi}\ & =\ & \frac{2(b-r)^2}{r^2(b+r)} \frac{1+r^2}{\sqrt{4+(b-r)^2}}\nonumber\\
    n\ & =\ & \frac{4br}{(b+r)^2}\nonumber\\
    k\ & =\ & \sqrt{\frac{4n}{4+(b-r)^2}},
\end{eqnarray}
where, the impact parameter 
\begin{equation}
    b = \frac{a \cos(i)}{R_{E}}  \frac{1-e^{2}}{1+e \cos(f-\omega)}.
\end{equation}
Here,  
\begin{equation}
    f = \tan^{-1} \left(\frac{1}{\tan\ \Omega\ \cos\ i}\right), 
\end{equation}
and $r$ is the ratio of the LC radius and the BH's Einstein radius, $R_{\rm{LC}}/R_{\rm{E}}$. The Einstein radius 
\begin{equation}
    R_{\rm{E}}=\sqrt{\frac{4GM_{\rm{BH}}}{c^2} \frac{D_{\rm{LS}} D_{\rm{L}}}{D_{\rm{S}}}}
\end{equation}
% %
where, $D_{\rm{LS}}$, $D_{\rm{L}}$, and $D_{\rm{S}}$ are the source-lens, lens-observer, and source-observer distances, respectively. For SL, $D_{\rm{L}}\approx D_{\rm{S}}$ and $D_{\rm{LS}}$ is given as;
\begin{equation}
    D_{\rm{LS}} = a \sin (i) \frac{1-e^{2}}{1+e \cos(f-\omega)}.
\end{equation}
The SNR of the SL signal is defined as
\begin{equation}
    \frac{\Delta \mathcal{F}}{\langle\mathcal{F}\rangle} = \mu_{\rm{SL}}-1
\end{equation}
where $\mu_{SL}$ is the magnification factor by which brightness of LC is modified during eclipses.

\subsection{Detection criteria}
\label{S:detect-criteria}
We employ three necessary conditions to determine the detectability of a particular detached BH--LC binary through photometric variations either using \gaia\ or \tess: 
\begin{enumerate}
    \item $\snr\geq10$. 
    \item The average apparent magnitude of the LC {\em G} ($m_{\rm{LC}})\leq20$ ($25$), the limiting magnitudes for \gaia\ (\tess). 
    \item At least one full orbit of the BH--LC binary must be observed. For \tess, this condition can be expressed as $\porb\leq\tdur$, where $\tdur$ is the duration of observation given the Galactic coordinates of the source. 
    For \gaia\ we consider the full extended mission duration of 10 years by imposing $\porb/\yr\leq10$.  
\end{enumerate}
We call the subset of BH--LC binaries in each of our MW realisations that satisfy the above conditions as the `optimistic' set for detectable BH--LC binaries. Essentially, this is a collection of sources that are resolvable through photometric variations. However, even when the SNR for photometric variability is high enough to be resolved, false positives may come from several other sources of uncertainties \citep[][]{Brown_2003,Sullivan2015,Kane_2016,Simpson_2022}. To minimize the possibility of false positives, we impose an additional condition, $\mbh\ge\mlc$. We call the subset of BH--LC binaries satisfying this additional condition as the `pessimistic' set. This additional condition $\mbh\ge\mlc$ ensures that a potential candidate LC exhibiting the desired photometric variation not only is the dominant source of light, but also is lower mass compared to the dark companion. Note however, even this additional condition may not be enough to rule out all possibilities of impostor BHs, especially in case of post main sequence (PMS) companions. A prominent example would be algol-type binaries where PMS components are paired with more massive non-degenerate companions \citep[][]{Gomel_2023}. Hence, photometric variations alone are likely not adequate for confirmation of BH--LC binaries. These should be considered as candidates ripe for followup spectroscopic observations.
\section{Results}
\label{S:results}

In this section, we describe the key properties of the present-day simulated BH--LC populations and discuss the detectable populations from EV, \beam\ or SL. We have already discussed the intrinsic populations without any restrictions in Paper\ I. We encourage readers to refer to Paper\ I for an exhaustive discussion on the present-day intrinsic BH--LC properties in the MW. Here we highlight a few key properties for which binary interactions and the choice of supernova physics leave clear imprints directly influencing their detectability via photometric variations. A view to the intrinsic properties also helps illuminate the effects of selection biases in identifying the detectable populations. Throughout this work we focus only on detached BH--LC binaries with $\porb/\yr\leq3$ and $\porb/\yr\leq10$ keeping in mind the maximum observation durations of \tess\ and \gaia. The sizes of the \tess\ and \gaia-detectable BH--LC populations for each set of binary models and observational selection cuts are summarized in \autoref{tab:detection_Gaia} and \ref{tab:detection_TESS}.
\begin{figure*}
    \plotone{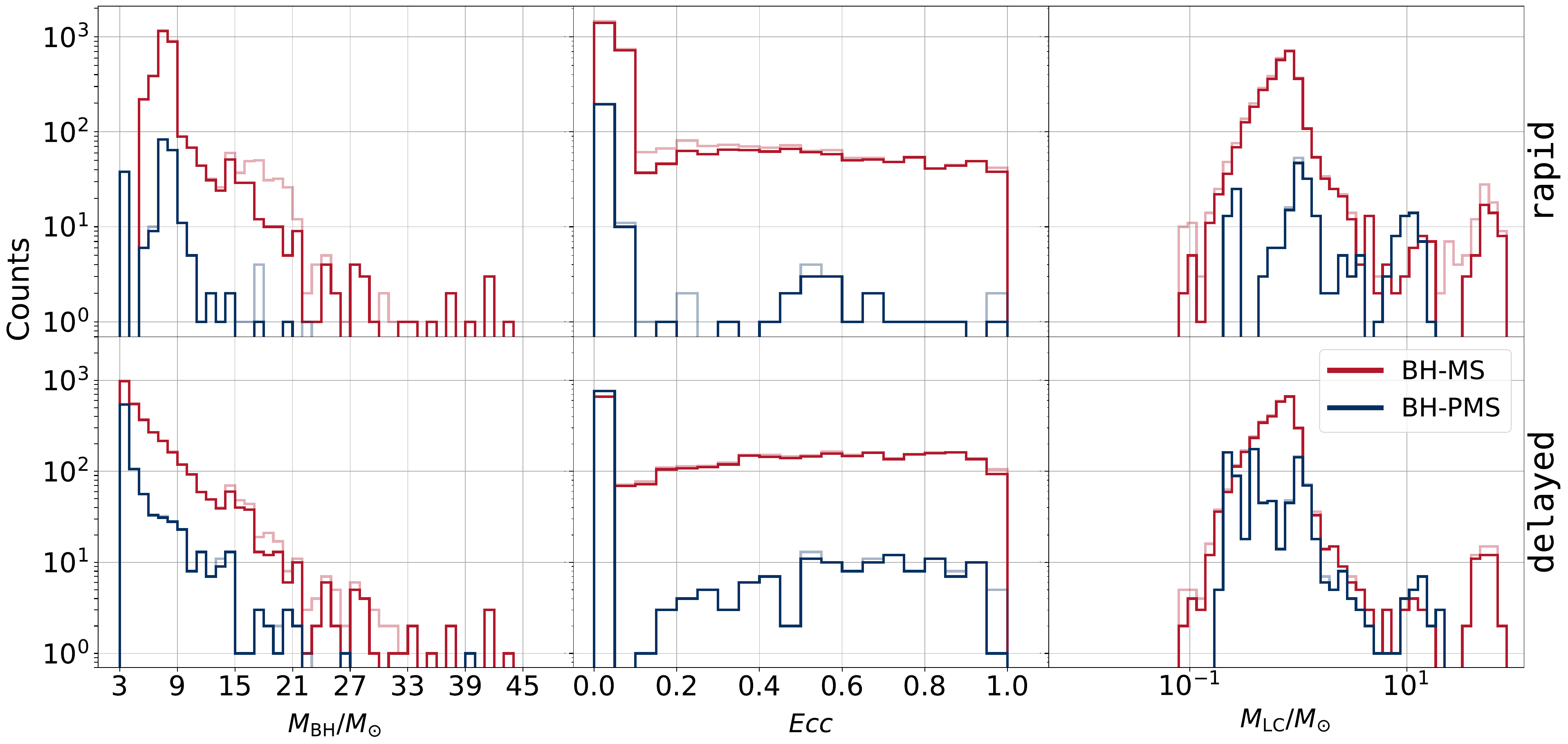}
     \caption{$\mbh$, $Ecc$ and $\mlc$ distributions of the present-day detached BH--LC binaries in the MW from our \rapid(top) and \delayed(bottom) models. The bright and faded curves represent detached BH--LCs with $\porb/\yr\le3$ and $10$, respectively. The red and blue curves represents BH--MS and BH--PMS,respectively. The distributions from the \rapid\ and \delayed\ models show significant differences between $\mbh/\msun=3$--$5$. BHs in this range come only from AIC of NSs in the \rapid\ model, whereas, the \delayed\ model allows BH formation both via AIC and CCSNe in this range.}
    \label{fig:intrinsic-pop-prop}
\end{figure*}

\begin{deluxetable*}{c|c|ccc|ccc}
\tablecolumns{8}
\tabletypesize{\small}
\tablecaption{BH--LC detectable population in the Milky Way by \gaia} 
\tablehead{
%
% \colhead{Model}\vline & \colhead{Type}\vline &
%  \multicolumn{6}{c}{Gaia}\\
% %
%  \cline{0-7}
 \colhead{Model}\vline & \colhead{Type}\vline & 
 \multicolumn{3}{c}{Optimistic}\vline & \multicolumn{3}{c}{Pessimistic($\mbh\ge\mlc$)}\\
\cline{3-8}
\colhead{}\vline & \colhead{}\vline  &
 \colhead{EV} & \colhead{\beam} & \colhead{SL}\vline & \colhead{EV} & \colhead{\beam} & \colhead{SL}
}
\startdata
 & MS & $358^{+24}_{-23}\ (765^{+34}_{29})$ & $772^{+42}_{-37}\ (1,193^{+48}_{-43})$ & $77^{+13}_{-11}\ (77^{+13}_{-11})$ &
  $333^{+25}_{-21}\ (678^{+34}_{-28})$ & $510^{+30}_{-28}\ (920^{+37}_{-36)}$ & $63^{+11}_{-9}\ (63^{+11}_{-9})$\\
  \rapid\ & PMS & $193^{+17}_{-21}\ (256^{+18}_{-20})$ & $173^{+16}_{-17}\ (305^{+20}_{-22})$ & $24^{+5}_{-6}\ (24^{+5}_{-6})$ &
  $160^{+16}_{-17}\ (211^{+16}_{-17})$ & $116^{+11}_{-14}\ (238^{+17}_{-18})$ & $24^{+5}_{-6}\ (24^{+5}_{-6})$ \\
  & total & $550^{+29}_{-29}\ (1,024^{+38}_{-34})$ & $942^{+46}_{-33}\ (1,501^{+46}_{-50})$ & $102^{+13}_{-13}\ (102^{+13}_{-13})$ &
  $493^{+25}_{-25}\ (889^{+35}_{-31})$ & $624^{+31}_{-28}\ (1,159^{+38}_{-38})$ & $86^{+13}_{-10}\ (86^{+13}_{-10})$ \\
  \hline
  & MS & $299^{+17}_{-24}\ (544^{+29}_{-26})$ & $355^{+26}_{-24}\ (659^{+33}_{-30})$ & $58^{+8}_{-8}\ (58^{+8}_{-8})$ &
  $252^{+18}_{-19}\ (459^{+27}_{-26})$ & $237^{+20}_{-18}\ (538^{+33}_{-28})$ & $44^{+7}_{-8}\ (44^{+7}_{-8})$ \\
  \delayed\ & PMS & $ 131^{+13}_{-14}\ (163^{+14}_{-17})$ & $62^{+9}_{-12}\ (189^{+19}_{-17})$ & $18^{+7}_{-5}\ (18^{+7}_{-5})$ &
  $127^{+13}_{-14}\ (157^{+12}_{-17})$ & $56^{+9}_{-9}\ (181^{+16}_{-18})$ & $18^{+7}_{-5}\ (18^{+7}_{-5})$\\
  & total & $428^{+28}_{-28}\ (707^{+35}_{-33})$ & $418^{+27}_{-26}\ (850^{+38}_{-37})$ & $76^{+11}_{-11}\ (76^{+11}_{-11})$ &
  $379^{+23}_{-27}\ (615^{+32}_{-32})$ & $294^{+23}_{-22}\ (719^{+35}_{-36})$ & $62^{+10}_{-10}\ (62^{+10}_{-10})$ \\
\enddata
 \tablecomments{Expected number of detached BH--LC binaries detectable by \gaia\ from EV, \beam\ and SL with SNR$\ge10\ (1)$ in the present-day Milky Way. The numbers and errors denote the median and the spread between the $10$th and $90$th percentiles across the Milky-Way realisations.}
\label{tab:detection_Gaia}
\end{deluxetable*}

\begin{deluxetable*}{c|c|ccc|ccc}
\tablecolumns{8}
\tabletypesize{\small}
\tablecaption{BH--LC detectable population in the Milky Way by TESS} 
\tablehead{
%
% \colhead{Model}\vline & \colhead{Type}\vline &
%  \multicolumn{6}{c}{Gaia}\\
% %
%  \cline{0-7}
 \colhead{Model}\vline & \colhead{Type}\vline & 
 \multicolumn{3}{c}{Optimistic}\vline & \multicolumn{3}{c}{Pessimistic($\mbh\ge\mlc$)}\\
\cline{3-8}
\colhead{}\vline & \colhead{}\vline  &
 \colhead{EV} & \colhead{\beam} & \colhead{SL}\vline & \colhead{EV} & \colhead{\beam} & \colhead{SL}
}
\startdata
 & MS & $46^{+10}_{-8}\ (176^{+16}_{-16})$ & $107^{+16}_{-12}\ (227^{+19}_{-22})$ & $1\ (3^{+3}_{-2})$ & $19^{+6}_{-5}\ (116^{+13}_{-12})$ & $34^{+8}_{-7}\ (149^{+15}_{-16})$ & $1\ (1^{+2}_{-1})$ \\
  \rapid\ & PMS & $38^{+9}_{-6}\ (78^{+13}_{-9})$ & $58^{+12}_{-9}\ (105^{+13}_{-11})$ & $0\ (1^{+2}_{-1})$ & $31^{+8}_{-6}\ (58^{+10}_{-9})$ & $18^{+6}_{-5}\ (47^{+8}_{-9})$ & $0\ (1^{+1}_{-1})$ \\
  & total & $86^{+12}_{-10}\ (253^{+22}_{-18})$ & $165^{+20}_{-16}\ (334^{+22}_{-23})$ & $1\ (4^{+3}_{-2})$ & $51^{+10}_{-9}\ (175^{+17}_{-16})$ & $52^{+10}_{-8}\ (196^{+14}_{-18})$ & $0\ (2^{+3}_{-1})$ \\
  \hline
  & MS & $42^{+10}_{-8}\ (117^{+14}_{-16})$ & $44^{+10}_{-7}\ (118^{+15}_{-14})$ & $0\ (0)$ & $17^{+6}_{-6}\ (75^{+13}_{-10})$ & $7^{+4}_{-3}\ (75^{+12}_{-12})$ & $0\ (0)$ \\
  \delayed\ & PMS & $34^{+8}_{-7}\ (67^{+10}_{-11})$ & $12^{+6}_{-4}\ (41^{+9}_{-9})$ & $0\ (0)$ & $32^{+8}_{-7}\ (62^{+10}_{-9})$ & $6^{+4}_{-2}\ (34^{+8}_{-8})$ & $0\ (0)$\\
  & total & $77^{+13}_{-11}\ (183^{+22}_{-17})$ & $57^{+10}_{-9}\ (159^{+17}_{-16})$ & $0\ (0)$ & $48^{+10}_{-9}\ (136^{+17}_{-13})$ & $14^{+5}_{-5}\ (109^{+14}_{-13})$ & $0\ (0)$ \\
\enddata
 \tablecomments{Same as \autoref{tab:detection_Gaia} but for TESS detectable BH--LC population.}
\label{tab:detection_TESS}
\end{deluxetable*}

\subsection{Intrinsic BH--LC population}
\label{S:intrinsic-prop-description}

\autoref{fig:intrinsic-pop-prop} shows the distributions of $\mbh$, $Ecc$, and $\mlc$ of the present day detached BH--LC populations for the \rapid\ and \delayed\ models with the relevant upper limits on $\porb$.
The characteristics of the intrinsic present-day populations depend strongly on the SNe explosion mechanism and the adopted binary evolution model which encodes mass transfer physics, stellar winds, and tidal evolution. However, we do not find significant differences between these distributions corresponding to $\porb/\yr\leq3$ and $\porb/\yr\leq10$. The $\mbh$ distribution spans the complete allowed $3-45\msun$ for both \rapid\ and \delayed\ models, however, striking difference is apparent near the so-called `lower mass-gap' between $3-5\msun$. While, both core-collapse SNe and AIC contribute in populating the BHs with $3\leq\mbh/\msun\leq5$ in the \delayed\ model, the BHs in this mass range in the \rapid\ model are produced from AIC only \citep[][]{Fryer2012}. As a result, the \rapid\ (\delayed) model consists $\sim1\%$ ($\sim55\%$) of BH--LC binaries with $\mbh/\msun\leq5$. Of the BH--LCs in the \delayed\ model with $\mbh$ in the lower mass-gap, $\sim10\%$ are produced from AIC and the rest are from CCSNe. 

The $Ecc$ distribution of the present-day population of BH--LC binaries transforms from the initially-assumed thermal distribution through binary stellar evolution including tides, mass loss, mass transfer, and natal kicks during BH formation. The \rapid\ and \delayed\ SNe prescriptions, through differences in the wind mass loss, birth mass function of BHs, and the details of the explosion mechanism, produce differences in the $\ecc$ distributions of present-day BH--LC binaries. We find that about $70\%$ of BH--LCs have near-circular ($Ecc\le0.1$) orbits in the \rapid\ model, in contrast to only $40\%$ of such systems in the \delayed\ model.

We find wide spreads in $\mlc$. For example, for most BHs with post main-sequence (PMS) or MS companions in detached orbits with $\porb\le3\yr$, $0.1\lesssim\mlc/\msun\lesssim20$. We find no significant difference between the $\mlc$ distributions for BH binaries with $\porb/\yr\le3$ and $\le10$. While we do find detached BH--PMS binaries with $20\lesssim\mlc/\msun\lesssim35$, they have $\porb/\yr\ge10$. Similarly, the majority of all BH--MS binaries with $\mlc/\msun\gtrsim18$ have $\porb/\yr\ge10$. Nevertheless, $\mlc$ in detached BH--MS binaries with short $\porb/\yr\le3$ exhibit a larger spread compared to that in BH--PMS binaries. About $2.1\%$ ($1.4\%$) of the present-day BH--LC population contains MS companions with $\mlc/\msun\ge35$ in the \rapid\ (\delayed)\ model. These are young binaries with ages $\le10\,\myr$ and are created from initially short-period ($\porb/\yr\le5$) binaries where the LC's progenitor got rejuvenated via mass accretion from the BH's progenitor via Roche-love overflow \citep[RLOF;][]{Tout1997}. Although only a small fraction of the overall population, these are potentially interesting systems. Over time, the high-mass LCs may evolve and create CO--CO binaries and be detected as a variety of interesting sources en-route. For example, during the later stage of the LC's evolution, the BH may start accreting via RLOF and become detectable via X-ray or radio emissions. Furthermore, if the mass transfer process is unstable, CE evolution may initiate and make the final CO--CO binary compact enough to emit detectable GWs or merge. Although interesting, following the final fate of these binaries is beyond the scope of this study.

\subsection{Gaia and TESS Detections}
\label{C:tess-detections}

\begin{figure*}
    \plotone{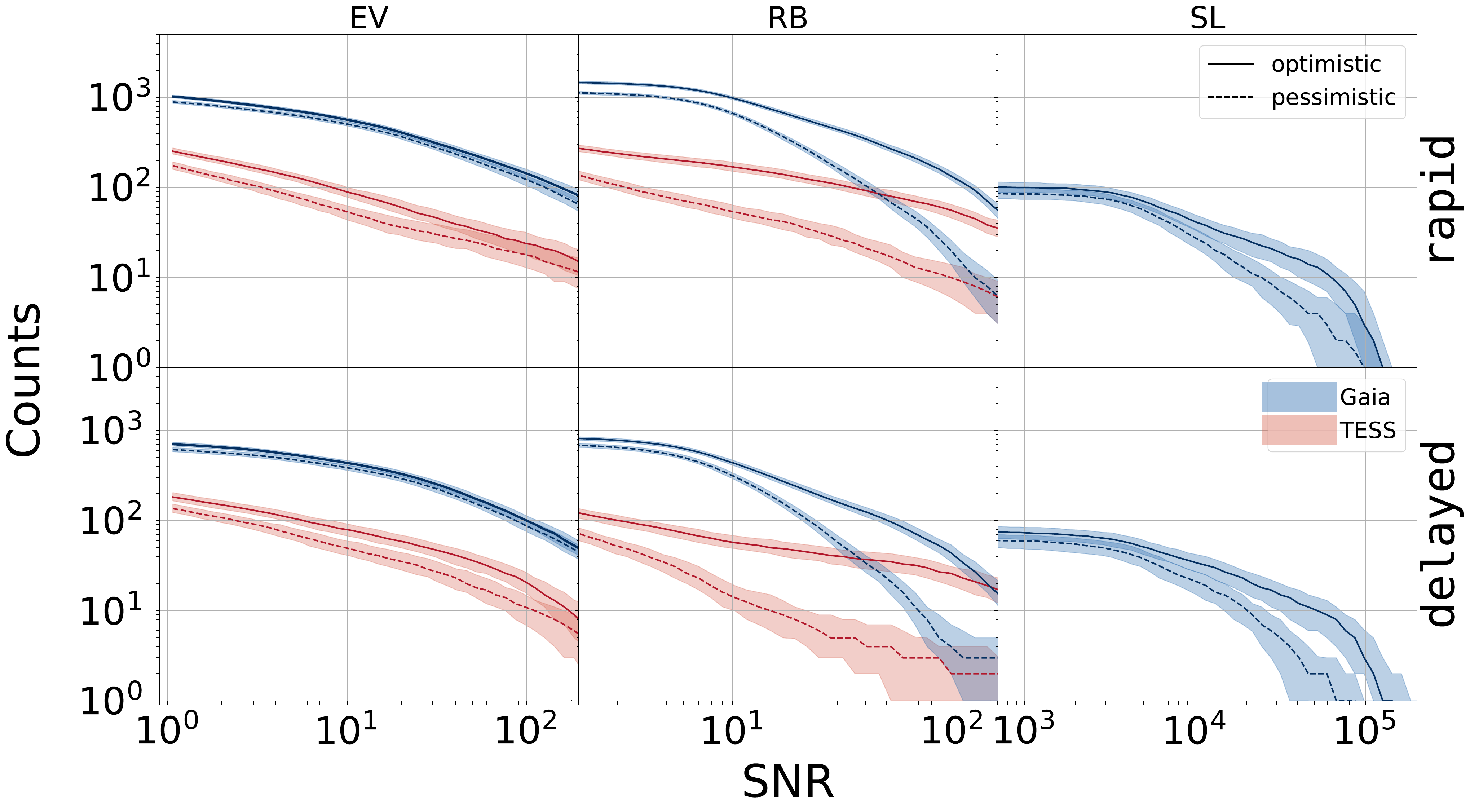}
    \caption{The reverse cumulative distribution of the expected detections of BH--LC binaries by EV (left) RB (middle) and SL (right) using \tess\ (red) and \gaia\ (blue) as a function of SNR for the \rapid\ (top) and \delayed\ (bottom) models. Solid and dashed lines represent the median number of detectable BH--LC binaries adopting the optimistic and pessimistic cuts (\autoref{S:detect-criteria}). The shaded regions represent the spread between the $10$th and the $90$th percentiles due to statistical fluctuations between our 200 independent MW realisations. }
    \label{fig:resolvable-numbers-ev}
\end{figure*}

\begin{figure*}
    \plotone{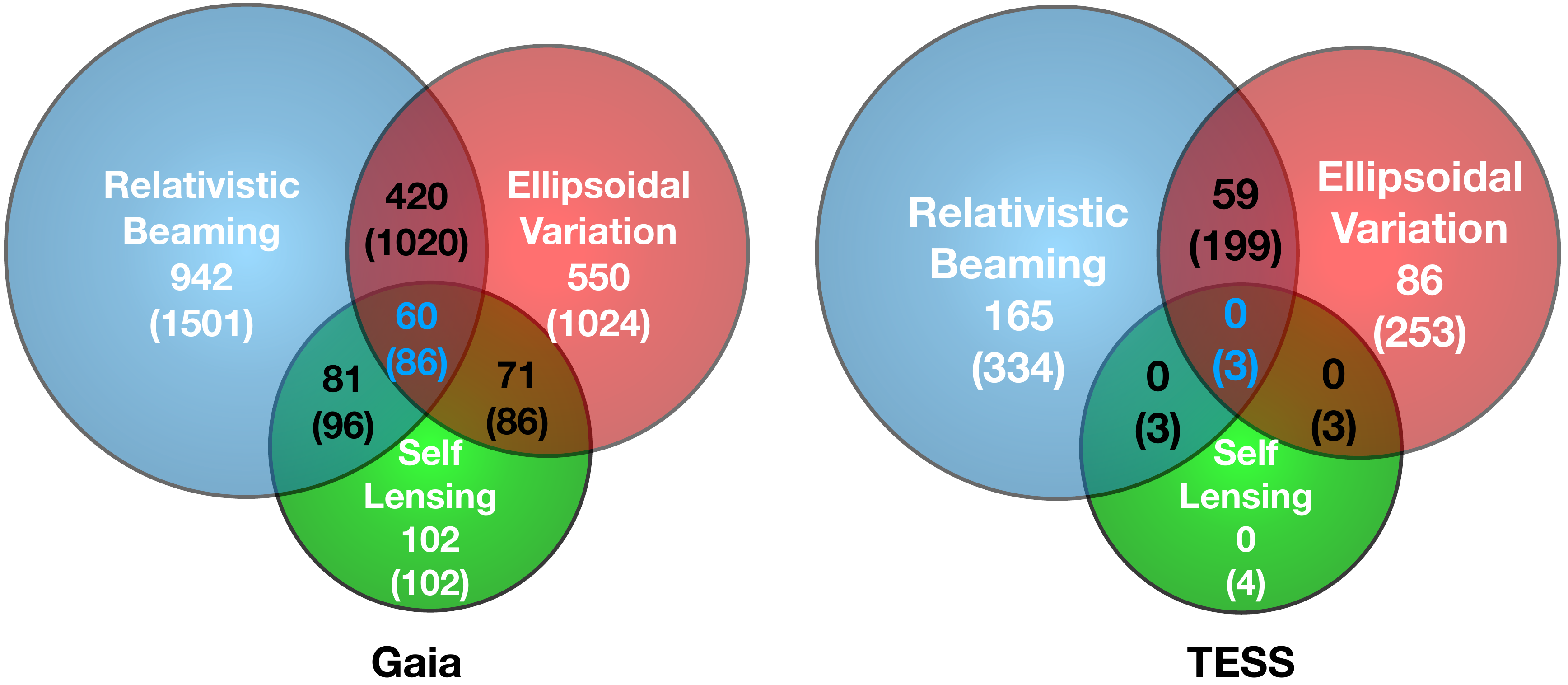}
    \caption{Number of \bhlc\ binaries resolvable via photometric variability by \gaia\ (left) and \tess\ (right) for the \rapid\ model with SNR $\ge10\ (1)$. Blue, red, and green denote populations resolvable via RB, EV, and SL signals, respectively. Numbers written in white, black, and blue denote total numbers for each set, number of systems with two and all three resolvable signals, respectively. For both telescopes, there are large ($\sim50\%$) overlaps between populations resolvable via EV and RB. The relatively small fraction of BH--LCs with detectable SL would also be detectable either via RB or EV or both. 
    (The equivalent figure (\autoref{fig:ev_rb_sl_overlap-delayed}) for our \delayed\ model is presented in the Appendix.)
    }
    \label{fig:ev_rb_sl_overlap}
\end{figure*}

\autoref{tab:detection_Gaia}, \ref{tab:detection_TESS} summarises the expected number of detections by \gaia\ and \tess\ via EV, RB, and SL with SNR$\ge10$ (1). In addition, we list the expected numbers where $\mbh\ge\mlc$. For both \gaia\ and \tess, photometric variations from EV and RB are significantly easier to detect compared to SL which leads to roughly an order of magnitude fewer detectable sources. This is expected for several reasons. In general, the geometric probability for SL is low, especially because here we consider only detached binaries which requires higher $\porb$. Even when the orientation allows SL, the magnification is typically low. Even if the maximum SL signal $\mu_{\rm{SL,max}}=[1+4/(R_{\rm{LC}}/R_{\rm{E}})^2]^{1/2}$ is greater than the photometric precision, the large impact parameter ($300-600 R_{\rm E}$) and short transit duration ($\sim 1$ hr) makes detection challenging with the cadence we have considered for \gaia\ and \tess. The overall low yield from SL is consistent with past studies \citep[][]{Rahavar_2010, Masuda2019,Wiktorowicz2021}.

Of course, in reality, an acceptable SNR is often dependent on the environment and the source itself. Hence, it is always instructive to investigate how the detectable numbers vary as a function of SNR. \autoref{fig:resolvable-numbers-ev} shows the reverse cumulative distribution of detached BH--LC binaries detectable via the various channels of photometric variations. For \rapid, in case of \gaia, the total number of detections using the optimistic cut with SNR$\ge10$ ($1$) is $1074$ ($1502$). The corresponding number using the pessimistic cut is $751$ ($1162$). Similarly, for \tess, the expected numbers of detections with SNR$\ge10$ and $1$ for the optimistic (pessimistic) cut are $194$ and $387$ ($80$ and $250$). 

Contribution from EV and RB are typically close to each other for both \gaia\ and \tess. Detected systems with EV and RB also have a high overlap (\autoref{fig:ev_rb_sl_overlap}); detectable detached BH--LC binaries with EV and RB have an overlap of roughly $50\%$ ($67\%$) for \tess\ (\gaia). In both \gaia\ and \tess, almost all detectable systems via SL can also be detected via either RB or EV or both. Overall, the expected number of detections in the \rapid\ model is higher by a factor of $\approx2$ compared to the \delayed\ model. This is simply because the \rapid\ model contains a higher proportion of higher-mass BHs compared to the \delayed\ model \citep[e.g.,][]{Fryer2012}. 

Note that, in our models, we adopt a very conservative lowest mass ($\mbh/\msun>3$) for BHs. Thus, these simulated numbers are for expected detectable BH--LC binaries with at least $\mbh/\msun>3$. This should reduce the possibility that the unseen object is a white dwarf or NS \citep[][]{Fonseca_2021,Romani_2022}. Moreover, our additional condition used in the pessimistic cut, $\mbh/\mlc\ge1$, should reduce false positives even further. Nevertheless, the confidence in identifying the nature of the dark component ultimately would depend on the estimated errors in the mass and followup observations \citep[e.g.,][]{Ganguly_2023,Chakrabarti_2023,El-badry_2022e,Shahaf_2023}. We envisage that while photometric variability can identify the interesting targets, multi-wavelength followup observations and RV followup will help to clearly identify the nature of the dark component in these binaries. 

\begin{figure}
    \plotone{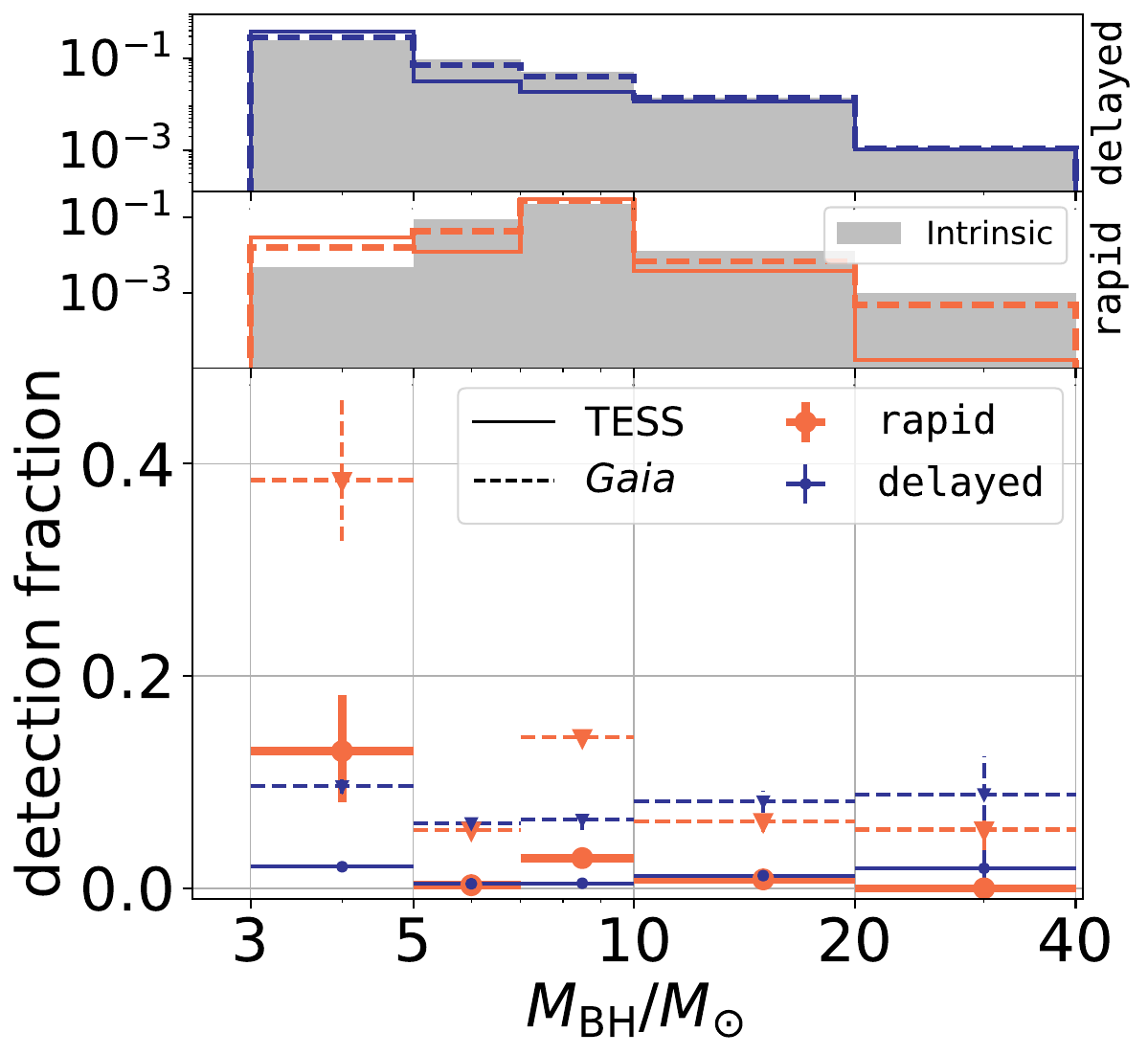}
    \caption{The ratio between the detectable binaries and intrinsic population (detection fraction) as a function of $\mbh$ for the BH--LCs detectable via photometric variations using \tess\ (solid) and \gaia\ (dashed) for the \rapid\ (orange) and \delayed\ (blue) models. The vertical and horizontal error bars represent the $10$--$90$th percentiles and the bin size in $\mbh$, respectively. The detection fraction is not strongly dependent on $\mbh$ for photometrically detectable BH--LC populations from both SNe models. The gray histogram in the top panel represents the intrinsic $\mbh$ distribution.
    \autoref{fig:mbh-multi-snr} illustrates varying the SNR threshold does not significantly affect the overlap between the intrinsic and observed populations.
    }
    \label{fig:detection-frac-vs-mbh-ev}
\end{figure}

Interestingly, similar to the case of astrometrically detectable BH--LC binaries presented in Paper\ I, we find that the photometrically detectable BH--LC binaries also show little dependence on the BH mass (\autoref{fig:detection-frac-vs-mbh-ev}, \ref{fig:mbh-multi-snr}). This can be somewhat counter-intuitive since the strength of the signal increases with increasing $\mbh$ for all physical effects we have considered here (see \autoref{S:transit-snr}). This is because, the detectability more strongly depends on the photometric precision of \gaia\ and \tess\ compared to the signal strength. The photometric precision, on the other hand, depends strongly on the magnitude of the LC \citep[][]{Rimoldini_2023} and does not depend at all on $\mbh$. As a result, the $\mbh$ distribution of the detectable population is expected to closely resemble the intrinsic one. This is in contrast to BH populations detected from other more traditional observations like X-ray, radio, and GWs \citep[][]{Jonker_2021,Liotine_2023}. 

\subsection{Key properties of the BH--LCs detectable via photometric variations}
\label{sec:key_properties}
\begin{figure*}
    \plotone{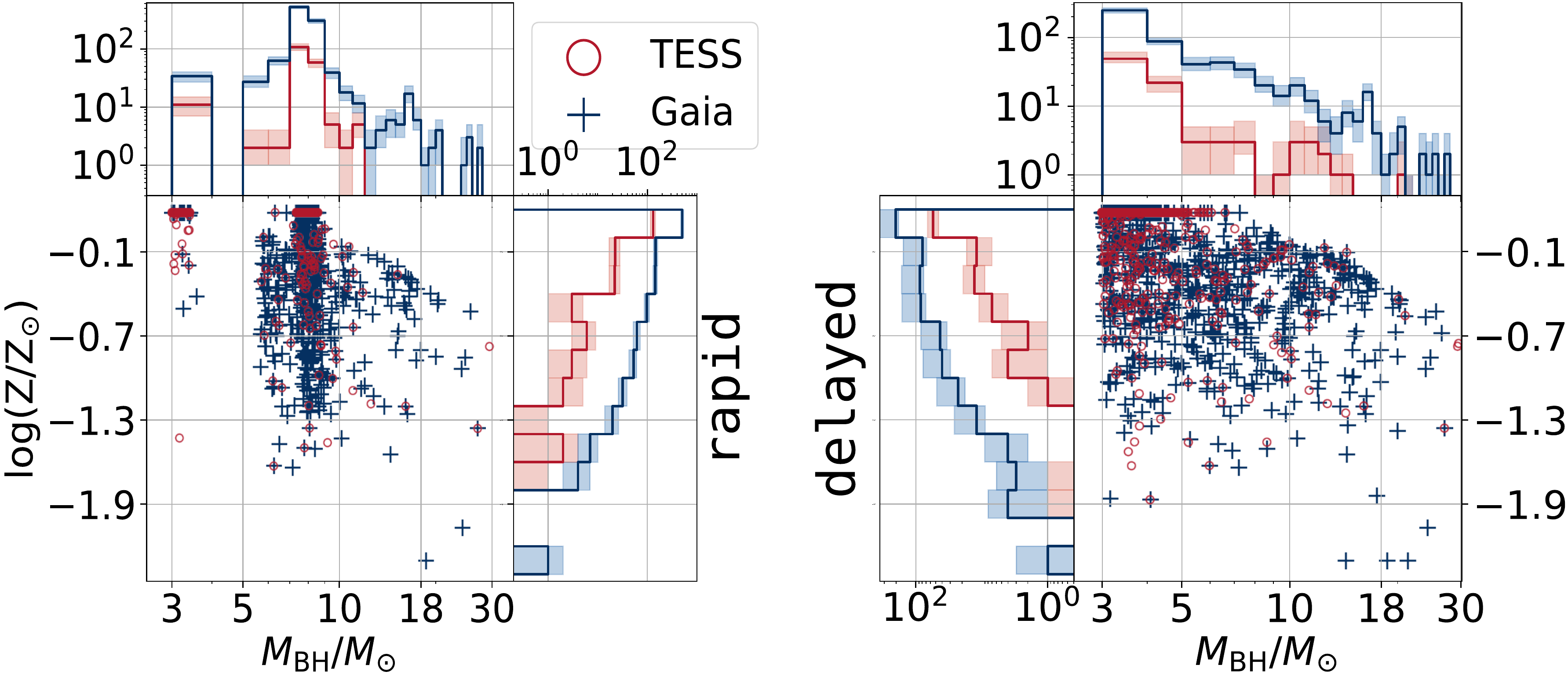}
    \caption{Distributions of $\mbh$ and progenitor metallicity of the detached BH--LC binaries detectable through photometric variability. Red circle and blue plus represent populations detectable using \tess\ and \gaia, respectively. Lines and shades in the histograms represent median and the spread between the 10th and 90th percentiles in each bin. Left and right figures denote the \rapid\ and \delayed\ models. }
    \label{fig:gaia_tess_mbh_met}
\end{figure*}
\begin{figure}
    \plotone{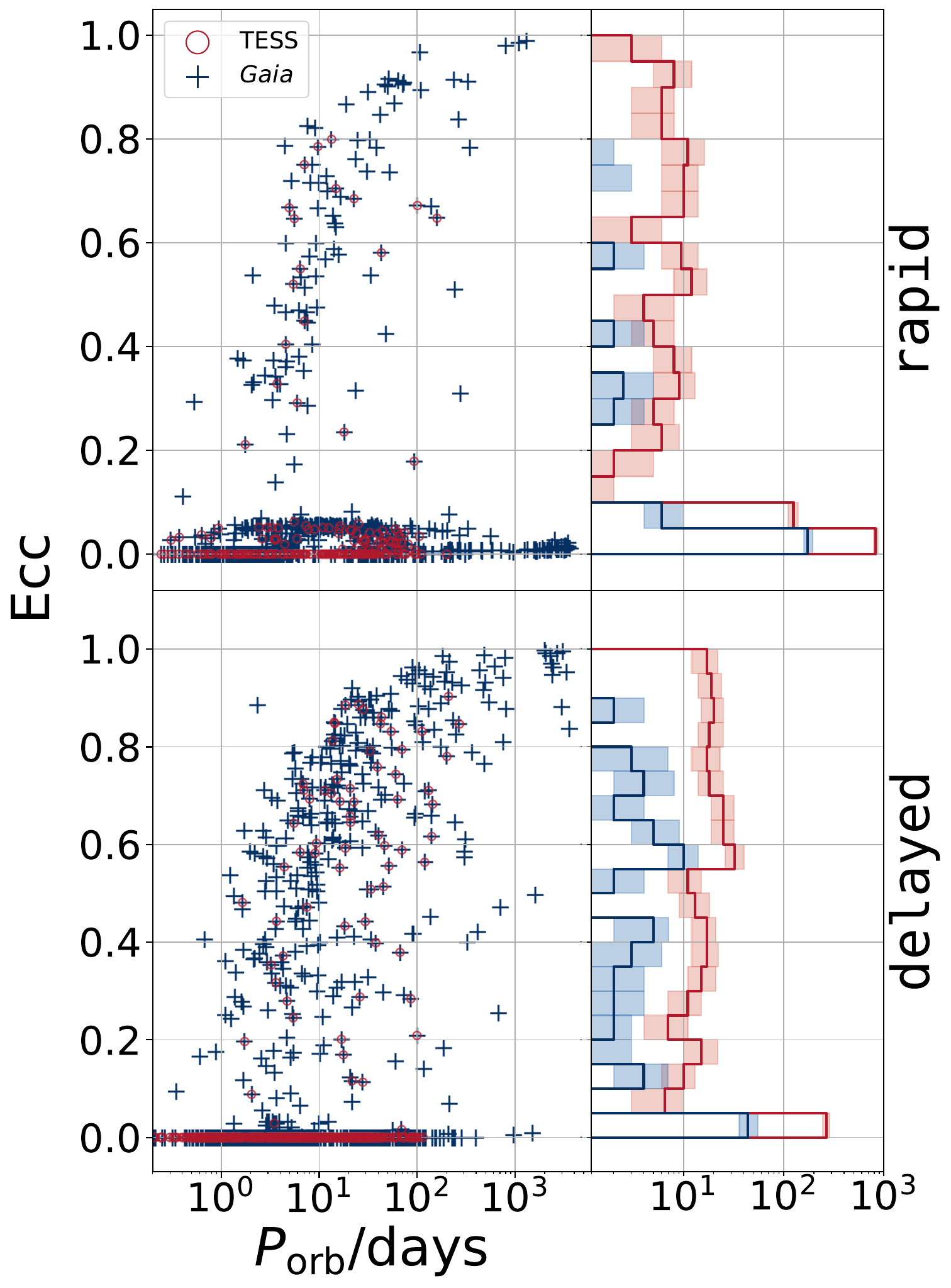}
    \caption{$\porb$ vs $\ecc$ for BH--LCs detectable via photometric variations using \tess\ (red circle) and \gaia\ (blue plus) for our \rapid\ (top) and \delayed\ (bottom) models. }
    \label{fig:gaia_tess_porb_ecc}
\end{figure}

Overall, the distributions of key observable properties for the detectable population are very similar to those of the intrinsic population. Moreover, the \tess\ and \gaia-detectable populations are very similar in properties. Detectable differences in the population properties come from the differences in the adopted supernova prescription. We use SNR$\ge10$ as the cutoff to determine the detectability, unless otherwise specified.

The distributions of the detected population through photometric variability show a wide spread in both \metal\ and $\mbh$ for both the \rapid\ and \delayed\ models (\autoref{fig:gaia_tess_mbh_met}). 
The $\mbh$ distribution shows distinct features for the \rapid\ and \delayed\ models. The lower mass gap ($3$--$5\,\msun$) in the intrinsic population for the \rapid\ model remains apparent also in the detectable population; $\sim6\%$ ($\sim3\%$) of the \tess\ (\gaia) detectable BH--LC binaries contain BHs with $3\leq \mbh/\msun\leq5$ in the \rapid\ model, in contrast to $\sim75\%$ ($\sim58\%$) in the \delayed\ model. Of course, in the \rapid\ model all detectable BHs in the mass gap must come from AIC of NSs. In contrast, in the \delayed\ model, most ($\sim 78-92\%$) detectable BHs in this mass range come from CCSNe and the rest from AIC. Contribution from AIC is a little higher ($22\%$ and $8\%$ for \tess\ and \gaia) in the \delayed\ model for detectable BHs with PMS companions. 

The metallicities of the detectable detached BH--LCs show a wide spread, $-2.4\leq\log(\metal/\metal\odot)\leq0.2$, although, the majority of the detectable population consist of young \bhlc's with $\metal\ge0.02$. As a result, we find $\mbh/\msun\le20$ in the detectable population. The wide spread in metallicities is particularly interesting. Using astrometric and photometric observations it may be possible to put constraints on the LC properties including metallicity and age. Based on these constraints, it may be possible to constrain the age and metallicity of the BH's progenitor in real systems. Furthermore, if the mass of the BHs can also be determined via photometric variations, astrometric solutions, or follow-up observations, then a metallicity-dependent map connecting BHs with their progenitors may emerge for a wide range in the observed metallicities. 

A metallicity-dependent map between progenitor properties and the BHs they create can be instrumental in improving our understanding of high-mass stellar evolution and binary interaction. These BH--LC systems descend from massive ($M>10\msun$) stars and go through several metallicity-dependent, important, but uncertain stages of evolution, such as mass-loss through winds, RLOF, and CE evolution. Thus, if indeed discovered in large numbers spanning a wide range in metallicities, the inferred mass of each component in the binary, combined with the metallicity and age of the LC, would help constrain models of the BH progenitor's evolution as well as high-mass binary stellar evolution via comparison between model and observed present-day properties of detached BH--LC binaries.

\begin{figure*}
    \plotone{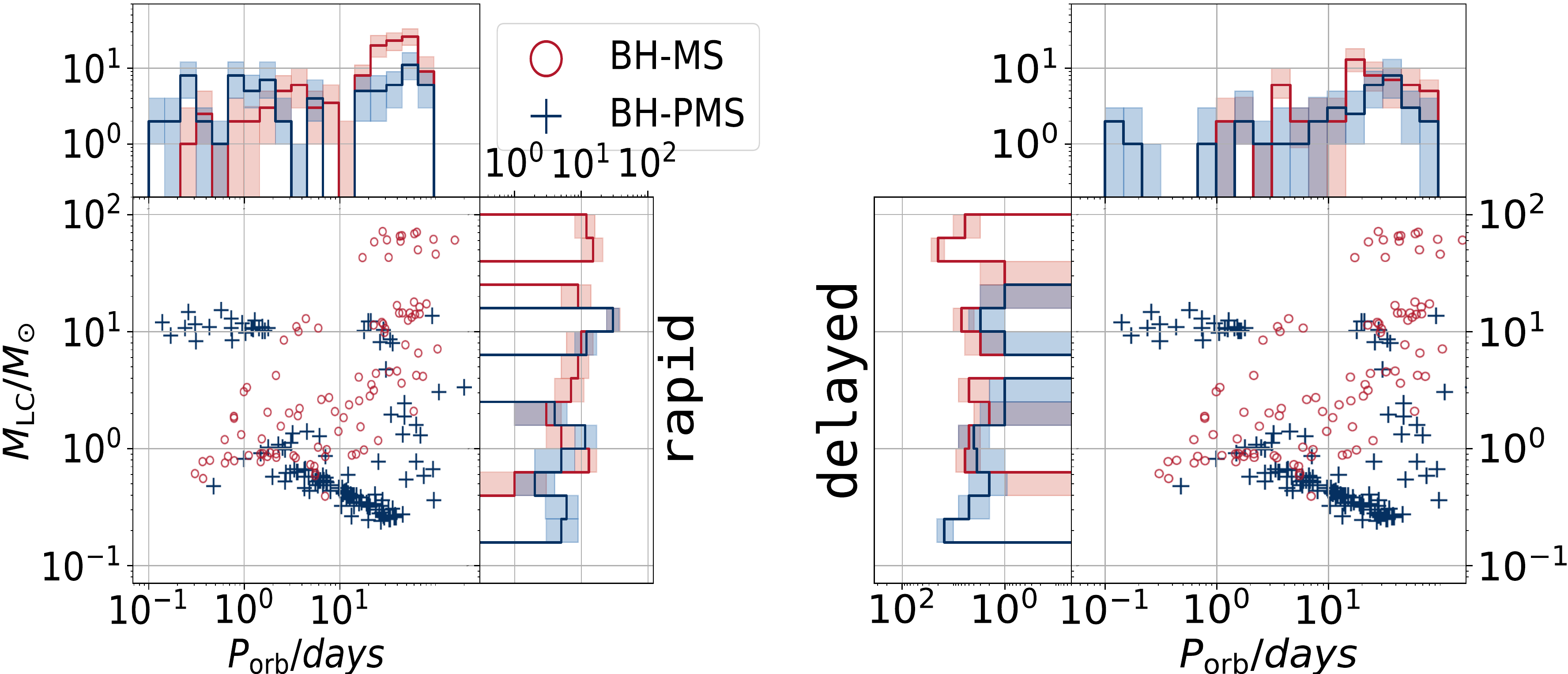}
    \caption{Distribution of $\porb$ vs $\mlc$ for detached BH--LCs detectable via photometric variability using \tess. Red circles and blue crosses represent BH--MS and BH--PMS binaries, respectively. Note the correlation between $\porb$ and $\mlc$, especially for the detectable BH--MS binaries.
    }
    \label{fig:tess_porb_mlc}
\end{figure*}

\begin{figure*}
    \plotone{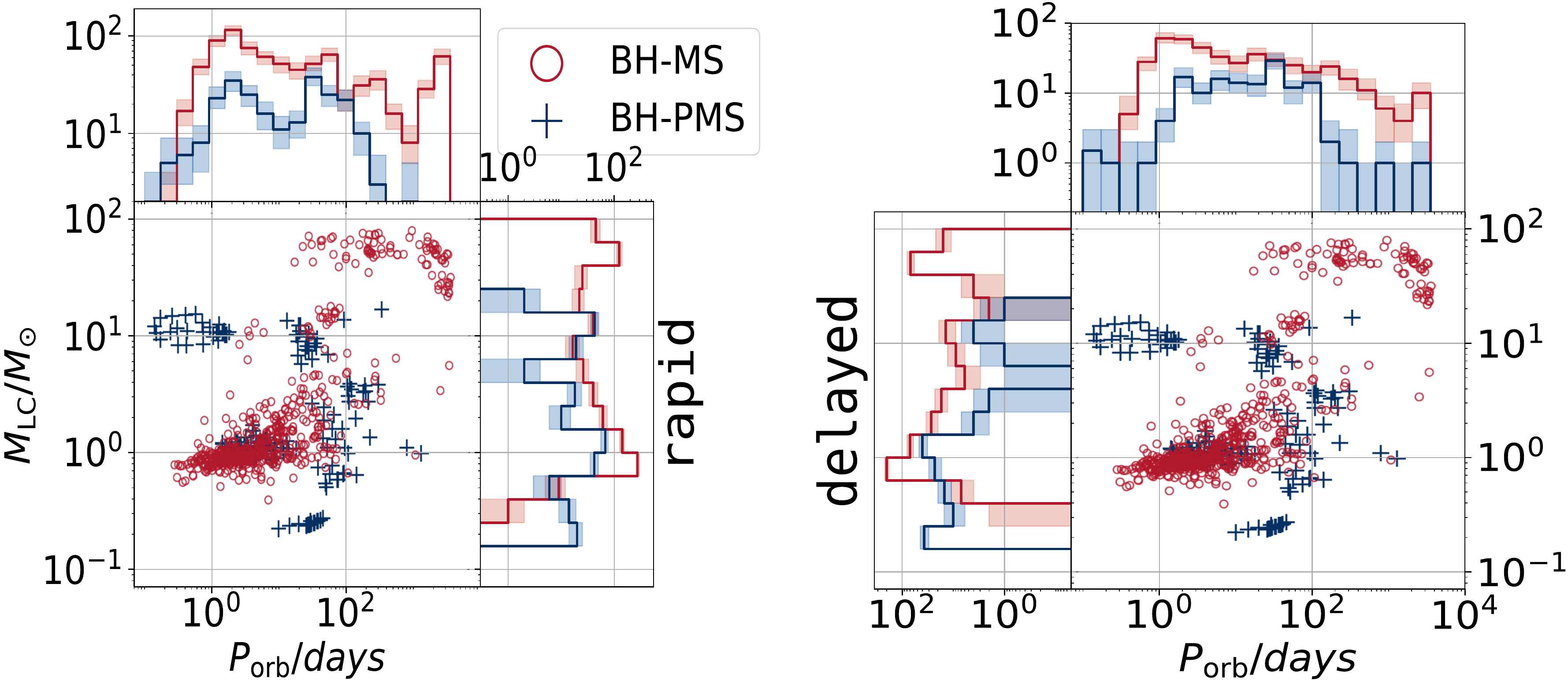}
    \caption{Same as \autoref{fig:tess_porb_mlc} but for BH--LCs detectable through photometric variability using \gaia. }
    \label{fig:gaia_porb_mlc}
\end{figure*}

Apart from the $\mbh$ distribution, the orbital eccentricities can also differentiate between different SN explosion mechanisms (also see Paper\ I). Majority ($\sim 60-98\%$) of the photometrically detectable BH--LCs in all our models go through at least one mass transfer or CE episode, which erases the initial orbital eccentricity. Thus, the final orbital eccentricity is almost entirely dependent on the natal kicks the BHs receive, which later can be further modified by tides depending on $\porb$ and the time since BH formation. Under the fallback-modulated prescription for natal kicks, the BHs in the \delayed\ model typically receive larger kicks compared to those in the \rapid\ model. As a result, the detectable BH--LC binaries in the \delayed\ model contain a much larger fraction ($\sim54\%$) of $\ecc>0.1$ orbits compared to those in the \rapid\ model ($6-11\%$). \autoref{fig:gaia_tess_porb_ecc} shows $\porb$ vs $\ecc$ for the detectable populations. The \rapid\ (\delayed) model contains about $94\%$ ($46\%$) and $89\%$ ($47\%$) BH--LC binaries with $Ecc\le0.1$ in the \tess\ and \gaia\ detected populations. Because of the relatively stronger natal kicks the BHs receive, the \delayed\ model contains a much higher fraction ($\sim30-35\%$) of BH--LC binaries with $\ecc>0.5$ compared to the \rapid\ model ($3-8\%$) in the \tess\ as well as \gaia\ detected populations. These observable differences in the $\ecc$ distributions can be really interesting if detached BH--LCs are indeed discovered in large numbers through photometric as well as astrometric channels. Since, the final orbital $\ecc$ is essentially dependent on natal kicks, a careful study of the $\ecc$ distribution for these systems 
should help in constraining poorly understood natal kick physics \citep[][]{Repetto_2017,Atri_2019, Andrews_2022b,Shikauchi_2023}.

We find that the $\porb$ distributions for the detectable BH--LCs exhibit diverse ranges extending up to $\sim100$ days for \tess\ and $\sim10$ years for \gaia, essentially limited by the observation duration (\autoref{fig:tess_porb_mlc}, \ref{fig:gaia_porb_mlc}), which also indicates that almost all detectable binaries have been observed for multiple transits. At first glance this is counter-intuitive because the signal is expected to be stronger for shorter $\porb$ for all channels of photometric variability (\autoref{eq:EV-lumin}, \ref{eq:RB-lumin}). This can be understood as a consequence of subtle effects from the formation channel of the BH--LC binaries detectable through photometric variations. Most ($\approx 45-71\%$ for \tess\ and $80-84\%$ for \gaia) detectable BH--LCs have gone through at least one CE episode during their evolution. The eventual detached configuration, for the majority of the detectable BH--LCs thus depends on when the CE ends. All else kept fixed, a lower-mass LC would require a larger orbital decay before the CE can be ejected. This introduces a correlation between $\porb$ and $\mlc$ (\autoref{fig:tess_porb_mlc}, \ref{fig:gaia_porb_mlc}). A higher $\mlc$ means a brighter target, which in turn means lower photometric noise, all else kept fixed. Thus, a combination of population properties as well as selection biases effectively reduces the strong $\porb$ dependence of the signal strength in the detectable population. 

Overall, we find that CE evolution plays a major role in shaping the properties of the BH--LCs detectable through photometric variability. For BH--MS binaries in the \rapid\ and \delayed\ models, $77\%$ and $52\%$ ($75\%$ and $80\%$) of the \tess\ (\gaia) detectable populations go through at least one CE evolution. In case of BH--PMS binaries, between $10$ to $20\%$ of the detectable systems go through CE evolution more than once. The detectable BH--PMS binaries also show interesting clustering in the $\porb$ vs $\mlc$ plane. The short-$\porb$ group contains systems with significantly higher $\mlc$ compared to that with longer $\porb$. The more compact BH--PMS binaries initially had massive progenitors ($\mlc_{\rm{,ZAMS}}\ge 20\,\msun$) in tight orbits ($\porb\lesssim10^2$ days). For these binaries, the CE is initiated by mass transfer from the LC and at the time of observation, the LC is a stripped Helium star. These are all younger than $8\,\myr$ at the time of observation. Because of the prevalence of CE evolution in the detectable BH--LCs, the properties of the observed populations may be able to put meaningful constraints on the uncertain aspects of CE evolution \citep[][]{Ivanova2013,Hirai_2022,Renzo_2023}.  

\section{Combination of different detection channels}
\label{S:follow-up}

\begin{figure}
    \plotone{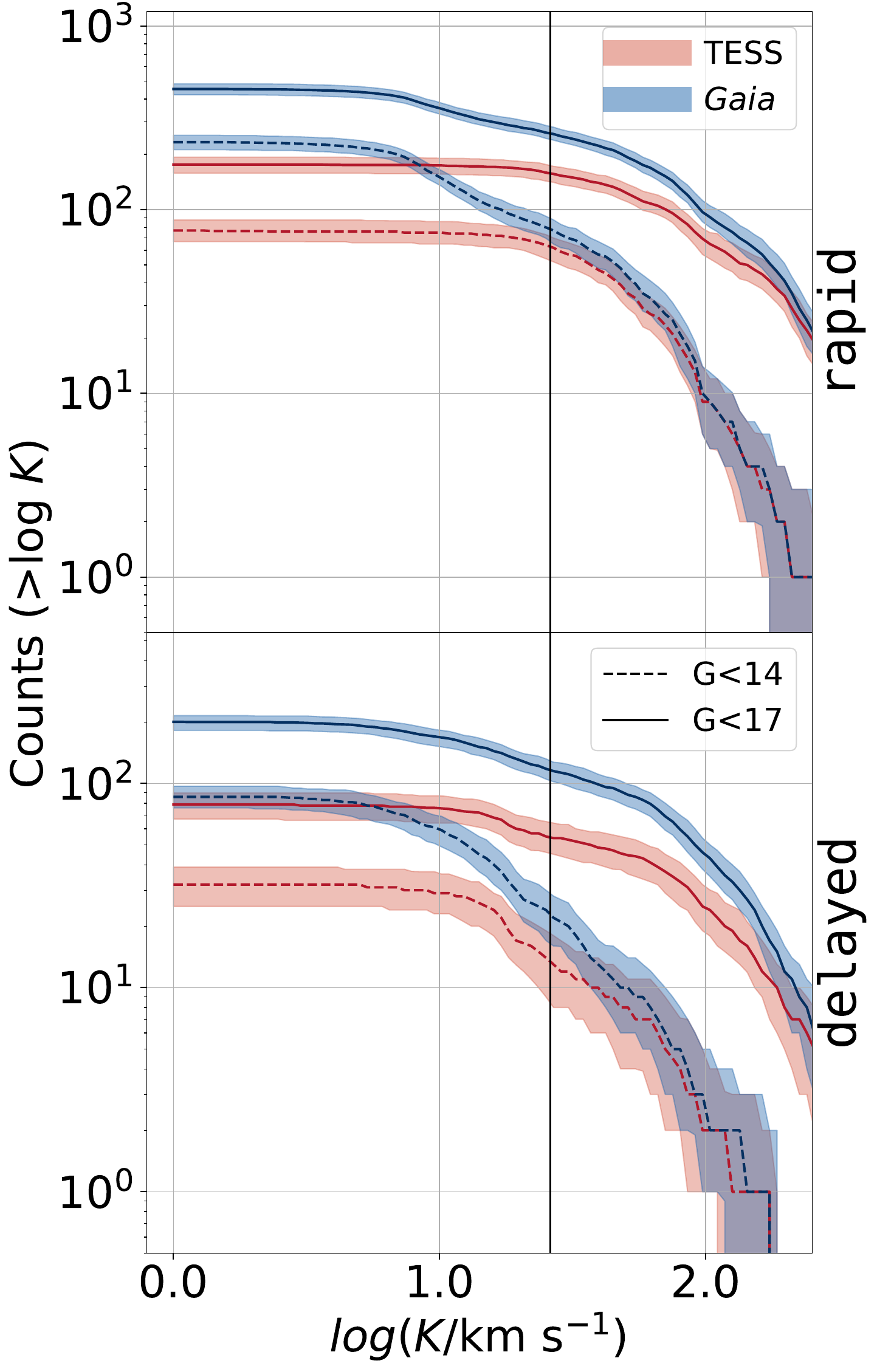}
    \caption{The reverse cumulative distribution of the RV semi-amplitude for detached BH--LCs resolvable via photometric variability using \gaia\ (blue) and \tess\ (red). The solid and dashed lines denote the BH--LC binaries with G $\le17$ and 14, respectively. Lines denote the median and the shaded regions denote the 10th and 90th percentiles from statistical fluctuations. Black vertical line denotes the minimum resolvable $K$ by \gaia.}
    \label{fig:rad-vel}
\end{figure}

Discovery of a population of stellar BHs in detached orbits with a LC is almost certainly going to receive a huge boost by combining various methods and followup studies. Indeed, several studies have identified candidate BH--LC binaries via various methods and combinations of them using \gaia's third data release \citep[DR3;][]{Andrews_2022a,Fu_2022,Gomel_2023,Jayasinghe_2023,Shahaf_2022,El-badry_2022e}. In Paper\ I, we highlighted that a large population of detached BH--LC binaries may be resolvable by \gaia's astrometry and that astrometry alone is likely to put strong enough constraints on the dark object's mass to clearly indicate a BH. Furthermore, we highlighted that \gaia's RV with a spectral resolution of $R\sim11,500$ for stars brighter than $G=17$ \citep[][]{Cropper_2018,Soubiran_2018,Sartoretti_2023}, itself could resolve the orbital motion for $\sim50-120$ astrometrically resolvable binaries depending on the model assumptions. 
Of course, once the candidates are identified, spectroscopic followup using higher-precision instruments can significantly improve these yields, but since \gaia's RV will automatically become available without any need for extensive followup, we only focus on that.

Based on \gaia's pre-mission estimates, the radial-velocity measurements were expected to be available for sources brighter than $G=17$ \citep[][]{Jordi2010}. However, with the DR3 release, the multi-epoch RV measurements are only available for sources with $G\le14$. Considering both end of mission expectation and the, likely more realistic, threshold obtained from DR3, we identify the photometrically resolvable (SNR$\ge10$ ) BH--LC populations with $G\le14$ ($17$), which would also be resolvable by \gaia's spectroscopy. \autoref{fig:rad-vel} shows the reverse cumulative distribution of the RV semi-amplitude for BH--LCs brighter than $G=14\ (17)$ and resolvable through photometric variability by TESS and \gaia. The vertical line shows \gaia's spectral resolution cutoff. We find that $62^{+9}_{-9}$ ($158^{+15}_{-17}$) and $14^{+10}_{-10}$( $55^{+10}_{-10}$ ) BH--LCs for $G\le14\ (17)$ in the \tess\ resolved population would also be resolved with the help of spectroscopy in the \rapid\ and \delayed\ models, respectively. Similarly, the numbers for spectroscopically resolved BH--LC binaries in case of \gaia\ photometry are $78^{+11}_{-11}$ ($260^{+24}_{-19}$) and $23^{+6}_{-6}$( $116^{+14}_{-12}$).  Interestingly, $25\%$--$60\%$ of all photometrically detectable BH--LC binaries brighter than $G=17$, ($15$--$30\%$ overall) are expected to have RV resolvable by \gaia. Thus, a combination of photometric detection and \gaia's RV analysis is expected to provide credence to these discoveries and allow better characterisation of orbital and stellar properties. 

\begin{figure}
    \plotone{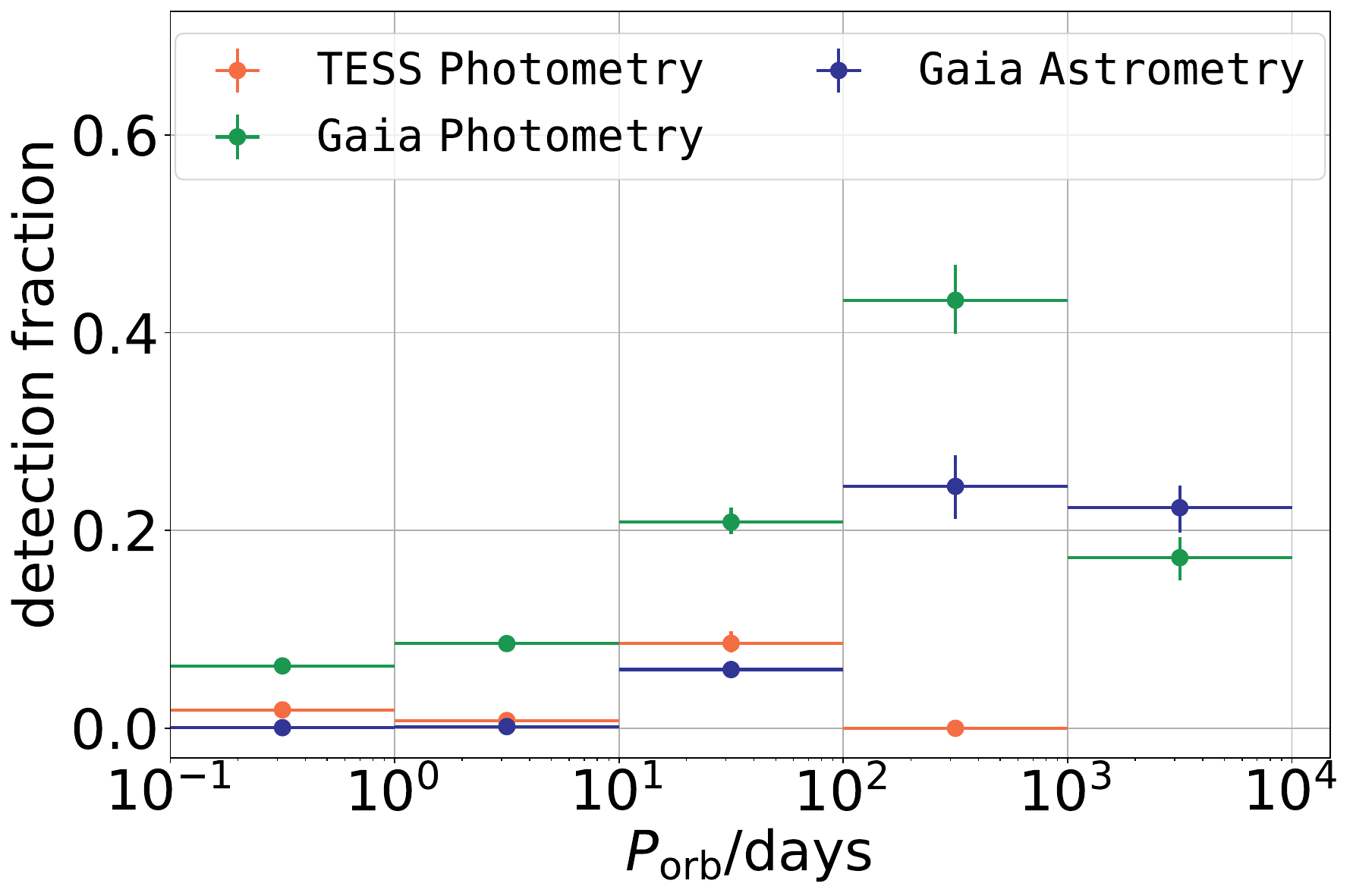}
    \caption{Detection fraction of detached BH--LC binaries as a function of $\porb$ via \gaia's astrometry (blue), \tess\ photometry (orange), and \gaia's photometry (green) for the \rapid\ model. Dots and error bars represent median, 10th, and 90th percentiles in each bin. Detection fraction via astrometry increases with increasing $\porb$ until it saturates for $\porb/\days\gtrsim10^2$. In contrast, detection fraction from photometry exhibits a peak. At large $\porb$, the decrease from the peak detection fraction is due to reduced photometric variability signal, whereas, at the small $\porb$, the decrease is due to the correlation between $\porb$ and $\mlc$ in the BH-LC binaries (\autoref{fig:tess_porb_mlc},\ref{fig:gaia_porb_mlc}).
    (The equivalent figure (\autoref{fig:detection-frac-vs-porb-delayed}) for our \delayed\ model is presented in the Appendix.)
    }
    \label{fig:detection-frac-vs-porb}
\end{figure}

\autoref{fig:detection-frac-vs-porb} shows the detection fraction as function of $\porb$ for detached BH--LCs detected via \tess\ and \gaia\ photometry and \gaia's astrometry. In case of \gaia's astrometry, the detection fraction monotonically increases until it saturates for $\porb\gtrsim100$ days. This of course is easy to understand; the larger the orbit, the easier it is to resolve via astrometry. The trend for the photometrically resolvable populations is more nuanced. In this case, both for \gaia\ and \tess, the detection fraction first increases with increasing $\porb$, peaks around $\porb/\days=10$--$100$ ($\porb/\days=100$--$1,000$) for \tess\ (\gaia) before decreasing. The peak is created due to the competition between two separate effects. The photometric variability signal depends strongly on $\porb$, the more compact the orbit, the stronger the signal. As a result, for sufficiently large $\porb$, the signal is simply too weak resulting in a decrease in the detection fraction. On the other hand, most detectable BH--LCs come from CE evolution. As a result, there is a distinct correlation between $\porb$ and $\mlc$ (\autoref{fig:tess_porb_mlc}, \ref{fig:gaia_porb_mlc}) and as a result, the magnitude. Hence, as $\porb$ increases, the photometric variability is easier to detect because of the lower noise for the brighter LCs. The different locations of the peaks for \gaia\ and \tess\ are reflective of their different observation duration. 

\begin{figure*}
    \plotone{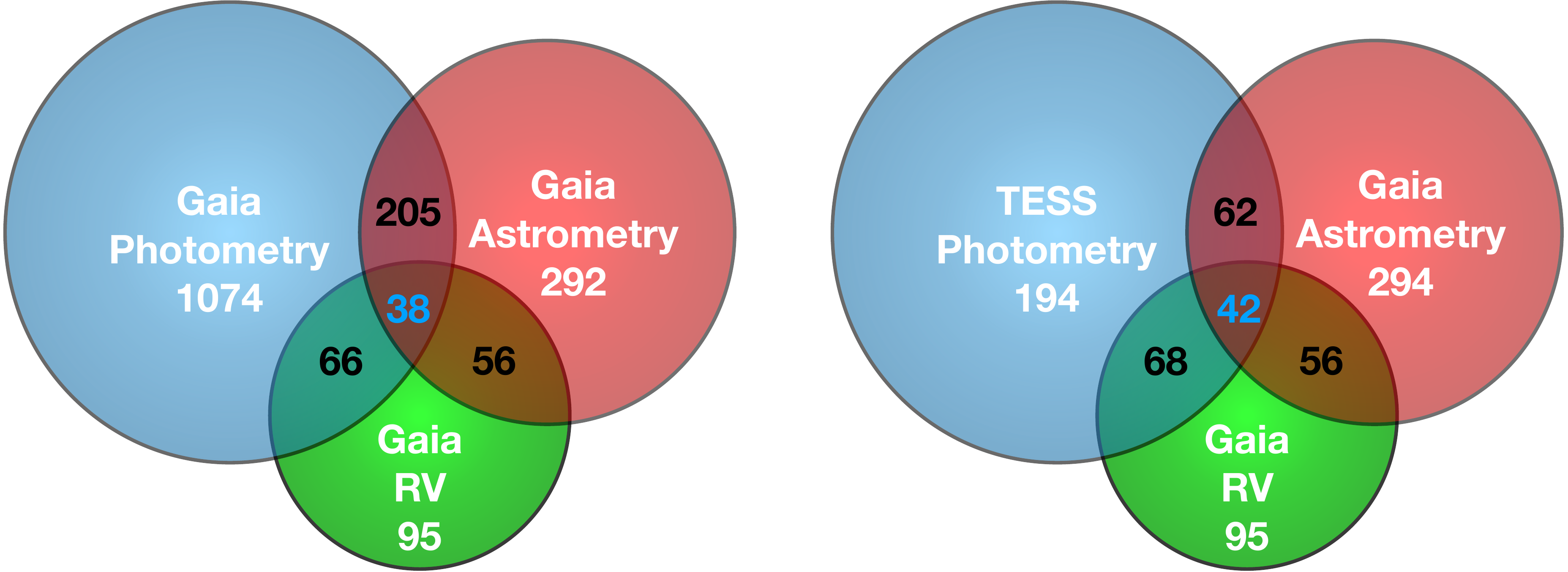}
    \caption{The expected numbers of detections from photomety (blue), \gaia's astrometry (red), and \gaia's RV (green) for detached BH--LC population for the \rapid\ model and their overlap. The white, black, and blue numbers denote yields in each set, overlap between two, and three different channels, respectively. 
    (The equivalent figure (\autoref{fig:gaia-rv-tess-obs-delayed}) for our \delayed\ model is presented in the Appendix.)
    }
    \label{fig:gaia-rv-tess-obs}
\end{figure*}

\autoref{fig:gaia-rv-tess-obs} shows the expected yields for detached BH--LCs from different detection channels and the overlap. We find between $10-30\%$ (depending on the adopted SNe model) of the photometrically detectable BH--LCs would also be resolvable via astrometry. On the other hand, between $5-30\%$ of the photometrically detectable BH--LCs are expected to have large enough RV to be resolved by \gaia's spectroscopy. Overall, about $5-20\%$ of all BH--LCs could be detectable from astrometry, photometry, as well as RV. 

Once the BH--LC candidates are identified considering an appropriate cut-off SNR and other resolvability criteria, the characterization  of the nature of the dark companion must be based on the $\mbh$ estimate from follow-up RV and astrometric measurements. Note that, the RV followup need not be limited to \gaia\ only. Furthermore, even in the absence of follow-up observations, light curve fitting of the photometric signal can be used to constrain the minimum $\mbh$ assuming edge-on orbit \cite[][]{Gomel_2021b,Rowan_2021}.

\section{Comparison with \gaia\ observed EV candidates}

\begin{figure}
    \plotone{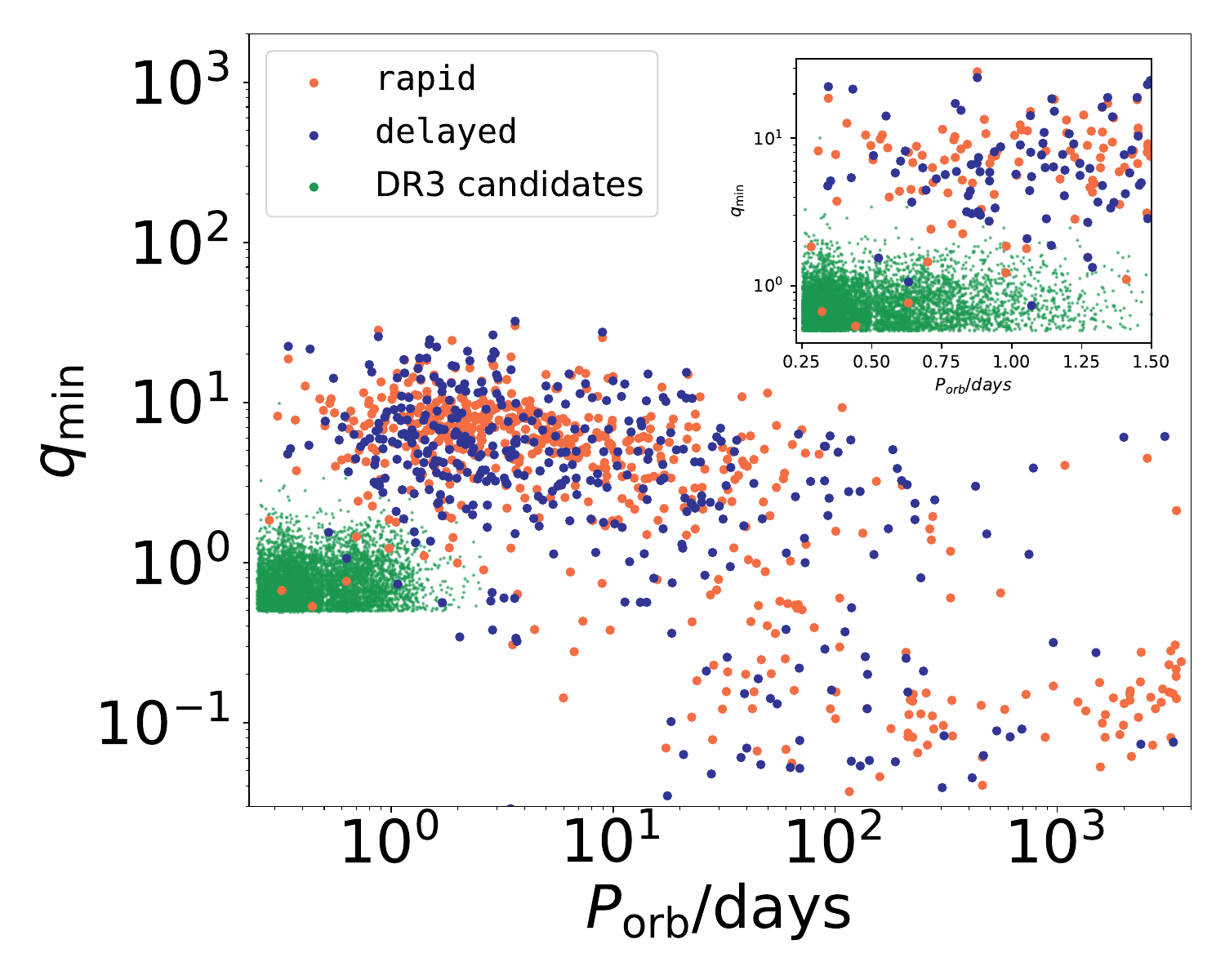}
    \caption{Mass ratio ($q$) vs $\porb$ distribution for \gaia\ detectable (SNR$\ge10$) simulated \bhlc\ binaries and \gaia\ DR3 detached CO--LC candidates from \citet[][]{Gomel_2023}. Orange and blue dots represent simulated BH--LC binaries in the \rapid\ and \delayed\ model, while, green dots represent the \gaia\ observed CO--LC candidates.} 
    \label{fig:q-vs-porb-ev}
\end{figure}

Using \gaia's third data release, \citet[][]{Gomel_2023} constructed a catalog of 6000 detached CO--LC candidate binaries identified using periodic flux variability. The study was predominantly focused on searching for CO--MS binaries in short-period ($\porb/\days\le2.5$) orbits. They avoided potential CO--PMS binaries because of a higher expected false-alarm rate \citep[][]{Gomel_2021b,Gomel_2023}. These candidates have $13\leq G\leq20$. Their inferred modified minimum mass-ratio $\qmin=\mco/\mlc$, is the mass-ratio assuming an edge-on orientation and a fillout factor of $0.95$ \citep[][]{Gomel_2021a}. Their \bhlc\ candidates have $0.5\leq q_{\rm{min}}\leq10$.

We perform mock observation of our model BH--MS binaries with $\snr>10$. While for the $\snr$ calculation, we generate orientations of our model BH--MS binaries, we assume edge-on configuration independent of these orientations, to calculate $\qmin$ for a direct comparison. \autoref{fig:q-vs-porb-ev} shows $\porb$ vs $\qmin$ for candidate observed \citep{Gomel_2023} and our model binaries. The model BH--LC binaries with $\porb/\yr\le10$ exhibit $\qmin$ between $0.01$ and $40$. We find a clear anti-correlation between $\qmin$ and $\porb$ stemming from the $\porb$-$\mlc$ correlation in BH--MS binaries arising from CE evolution (\autoref{sec:key_properties};\autoref{fig:tess_porb_mlc},\ref{fig:gaia_porb_mlc}). Binaries with low $\qmin$ contain rejuvenated massive LCs, where the LC progenitor have previously accreted mass from the BH progenitor through stable RLOF. We also find in our simulated population that the Porb  distribution has a peak around 2.5 days for the BH--MS population  and a shifted peak ~12 days for the BH--PMS population similar to that of \citet{Gomel_2021b}.  We find limited overlap between the observed candidates and our model binaries. This may be because most of the observed CO candidates may be WDs or NSs, and not BHs. On the other hand, in future if the number of photometrically detected BH candidates increases, and spans a wider range in $\porb$, we predict an anticorrelation.

\begin{figure}
    \plotone{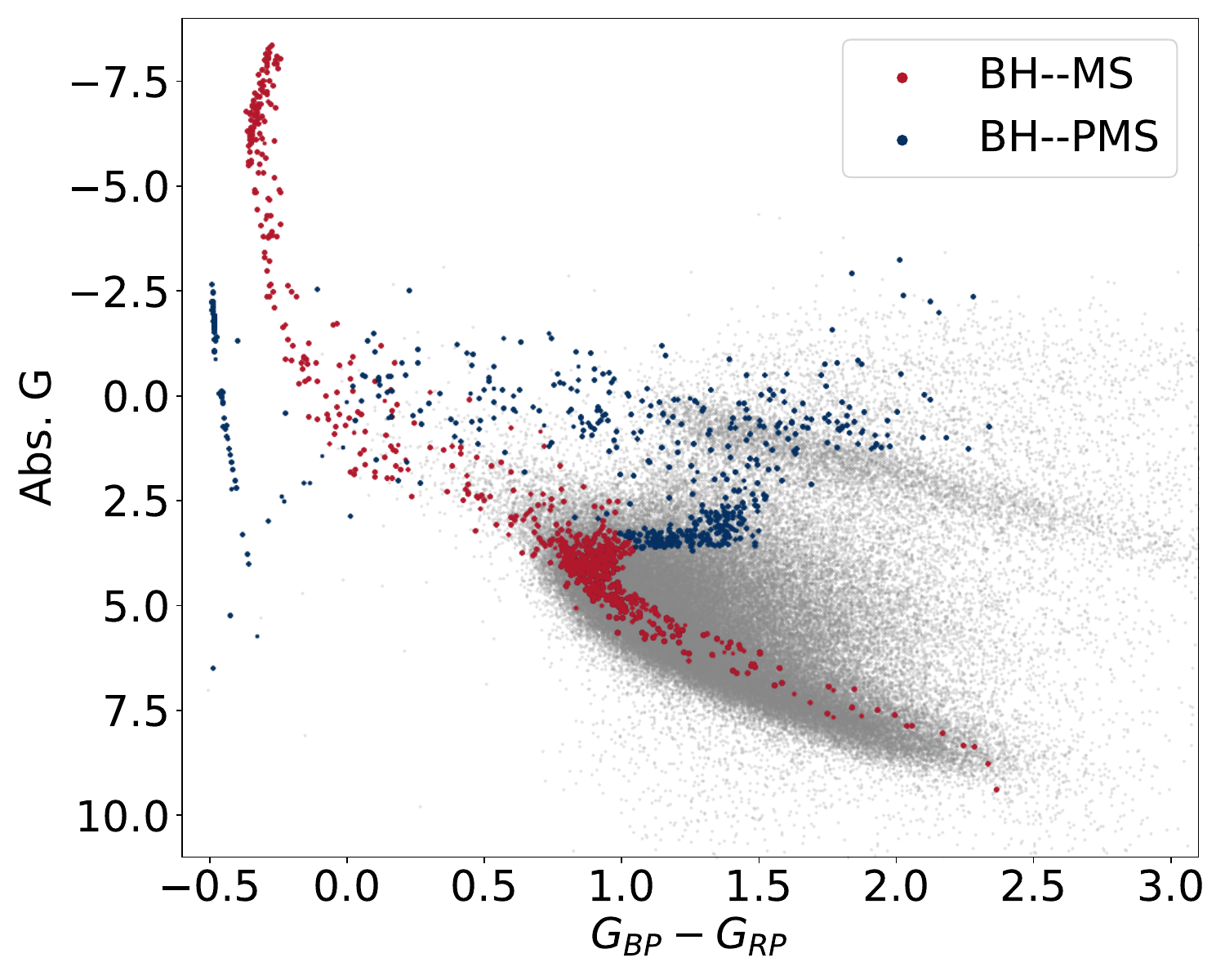}
    \caption{Distribution of BH--LC binaries in the \gaia's color magnitude diagram. The gray dots in the background represents stellar sources from \gaia\ archive, the simulated BH--MS and BH--PMS binaries are represented by red and blue dots, respectively.} 
    \label{fig:gaia-cmd}
\end{figure}

In \autoref{fig:gaia-cmd} we show our model detectable BH binaries (colored dots) with respect to \gaia's color-magnitude diagram (CMD), the grey dots represent all \gaia\ sources. The simulated BH--LCs align well with the MS and PMS giant regions on the CMD. The simulated resolvable BH--MS binaries brighter and bluer than the MS turn-off depict the rejuvenated sources. Overall, the CO--LC candidates identified from \gaia's DR3 illustrate that a potential population of CO--LC candidates exist and more candidates are expected to be found in future data releases.

\section{Summary and Discussions}
\label{S:conclusions}
We have explored the possibility of detecting detached BH--LC binaries via photometric variability with \tess\ and \gaia. We create highly realistic present-day BH--LC populations using the BPS suite $\cosmic$ \citep[][]{Breivik2020} taking into account a metallicity-dependent star formation history and the complex correlations between age, metallicity, and location of stars in the Milky Way \citep[][]{Wetzel2016,Hopkins2018,Sanderson2020}. We have used two widely adopted SNe explosion mechanisms, rapid and delayed \citep[][]{Fryer2012}, to create two separate populations of present-day BH--LC binaries. We have shown the key observable features of the intrinsic BH--LC populations adopting appropriate $\porb$ limits (see \autoref{S:intrinsic-prop-description}) as well as those that are expected to be detected via photometric variability (see  \autoref{C:tess-detections}). 

Using 200 realisations to take into account statistical fluctuations, taking into account different physical sources for photometric variability, \tess\ and \gaia\ selection biases, and three-dimensional extinction and reddening, we have generated a highly realistic population of detectable detached BH--LC binaries in the Milky Way at present (\autoref{S:synthetic-mw}, \ref{S:detect-criteria}). In addition to detection through photometric variability, we have also analysed \gaia's RV and astrometry to find relative yield and sources that could be detectable via multiple channels (see \autoref{S:follow-up}). 

\begin{itemize}
    \item We predict about $50-200$ and $300-1000$ detached BH--LC binaries may be detected by \tess\ and \gaia\ through photometric variability arising primarily from EV and RB.
    \item The photometrically detectable BH--LCs are expected to have wide range in metallicity and host BHs spanning a wide range in mass (see \autoref{fig:gaia_tess_mbh_met}). This is potentially interesting since in such systems, if the LC properties such as age and metallicity can be observationally constrained, we may be able to find a direct connection between the BHs and their progenitor properties. 
    
    \item The detection fraction is not strongly dependent on the BH mass (\autoref{fig:detection-frac-vs-mbh-ev}). Thus, the detectable BHs are expected to be similar in properties to the intrinsic BHs in detached BH--LC binaries. 
    
    \item The orbital $\ecc$ is essentially determined by BH natal kicks. As a result, if detected in large numbers, the $\ecc$ distribution can put constraints on natal kicks from core-collapse SNe. 
    
    \item Since a majority ($\sim 45-84\%$) of BH--LCs detectable through photometric variability using \tess\ and \gaia\ go through at least one CE episode, there is an interesting correlation between $\porb$ and $\mlc$, especially for BH--MS binaries (\autoref{fig:tess_porb_mlc}, \ref{fig:gaia_porb_mlc}). It will be interesting to verify this trend. Moreover, since this stems primarily from the energetics of envelope ejection, if detected in large numbers as we predict, this population may put meaningful constraints on the various uncertain aspects of CE physics. 
    \item A significant fraction of photometrically detectable BH--LC binaries may also be detectable via \gaia's RV and astrometry ($5$--$20\%$ are detectable via all three methods, \autoref{fig:rad-vel}, \autoref{fig:gaia-rv-tess-obs}), thus helping provide stronger constraints on their properties. 
\end{itemize}

In Paper\ I, we showed the potential of \gaia's astrometry for detecting and characterizing detached BH--LC binaries in large numbers. In this work we show that a combination of photometry, RV, and astrometry can significantly increase the number of identified detached BH--LC candidates. Especially, many detached BH--LCs are expected to be detectable via astrometry, RV, as well as photometry. Once identified, followup observations using more sophisticated instruments may improve the characterisation of these candidates even further. Our models suggest that we are on the verge of discovering a treasure trove in BH binaries, while the recent BH discoveries from Gaia astrometry \citep[][]{El-badry_2022e,El-badry_2023, Chakrabarti_2023} whet our enthusiasm.  

\begin{acknowledgements}
We thank the anonymous referee for insightful comments and constructive suggestions. CC acknowledges support from TIFR's graduate fellowship. SC acknowledges support from the Department of Atomic Energy, Government of India, under project no. 12-R\&D-TFR-5.02-0200 and RTI 4002. NS acknowledges TIFR's visiting summer research program during which this project was initiated. All simulations were done using cloud computing on Azure. The Flatiron Institute is supported by the Simons Foundation.
\end{acknowledgements}

\software{\texttt{Astropy}\ \citep{astropy:2013, astropy:2018, astropy:2022}; \cosmic\ \citep{Breivik2020}; \mwdust\ \citep{Bovy2016}; \texttt{isochrones}\ \citep{2015ascl.soft03010M}; \texttt{matplotlib}\ \citep{matplotlib}; \texttt{numpy}\ \citep{numpy}; \texttt{scipy}\ \citep{SciPy-NMeth_2020}; \texttt{ticgen}\ \citep[][]{ticgen_2017,Stassun_2018};\texttt{tess-point}\ \citep[][]{tess_point_2020};\texttt{pandas}\ \citep[][]{Pandas_2010};\texttt{sympy}\ \citep[][]{sympy_2017}}

\bibliographystyle{aasjournal}
\bibliography{tess_bh_lc}   

\appendix
\restartappendixnumbering

\section{Supplimentary figures}

\begin{figure}[h]
    \plotone{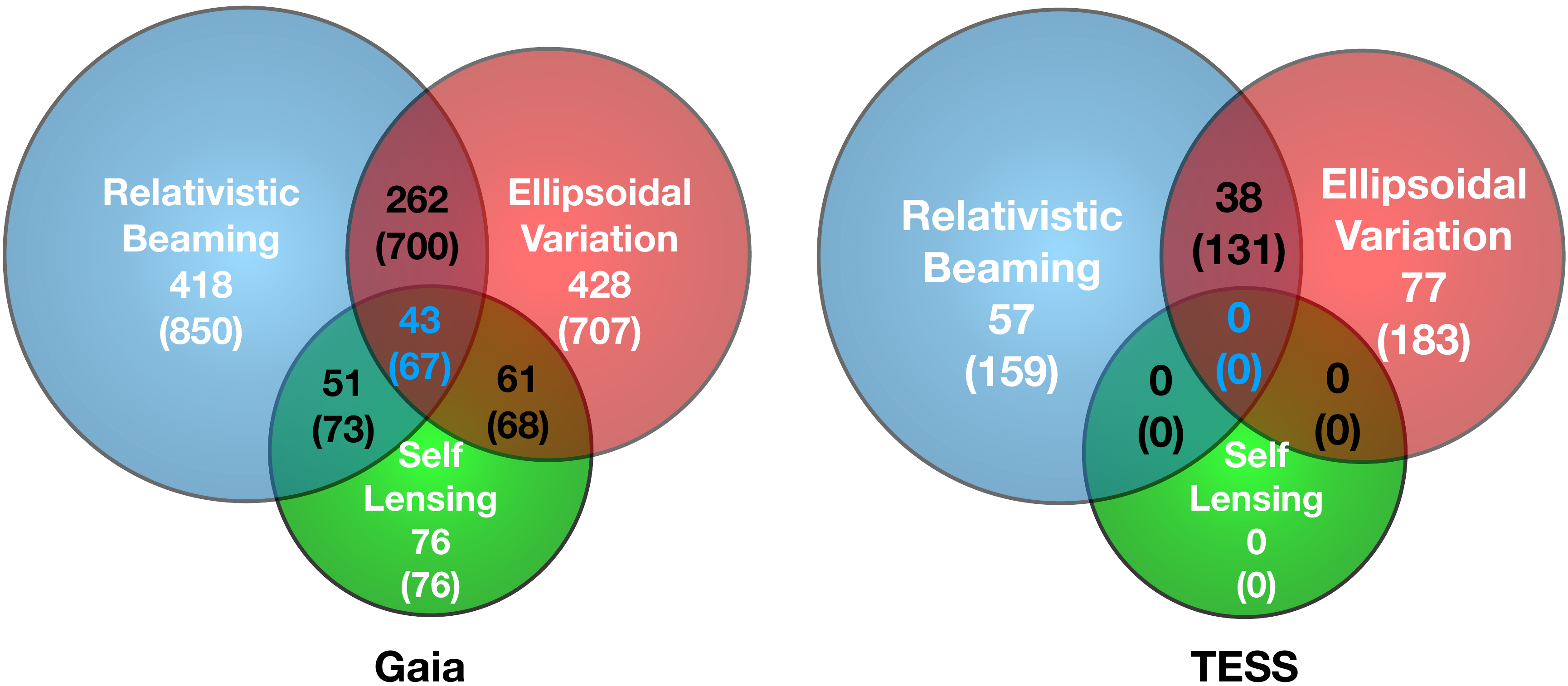}
    \caption{Same as \autoref{fig:ev_rb_sl_overlap} but for the BH-LCs in our \delayed\ model. 
    }
    \label{fig:ev_rb_sl_overlap-delayed}
\end{figure}

\begin{figure}[h]
    \plotone{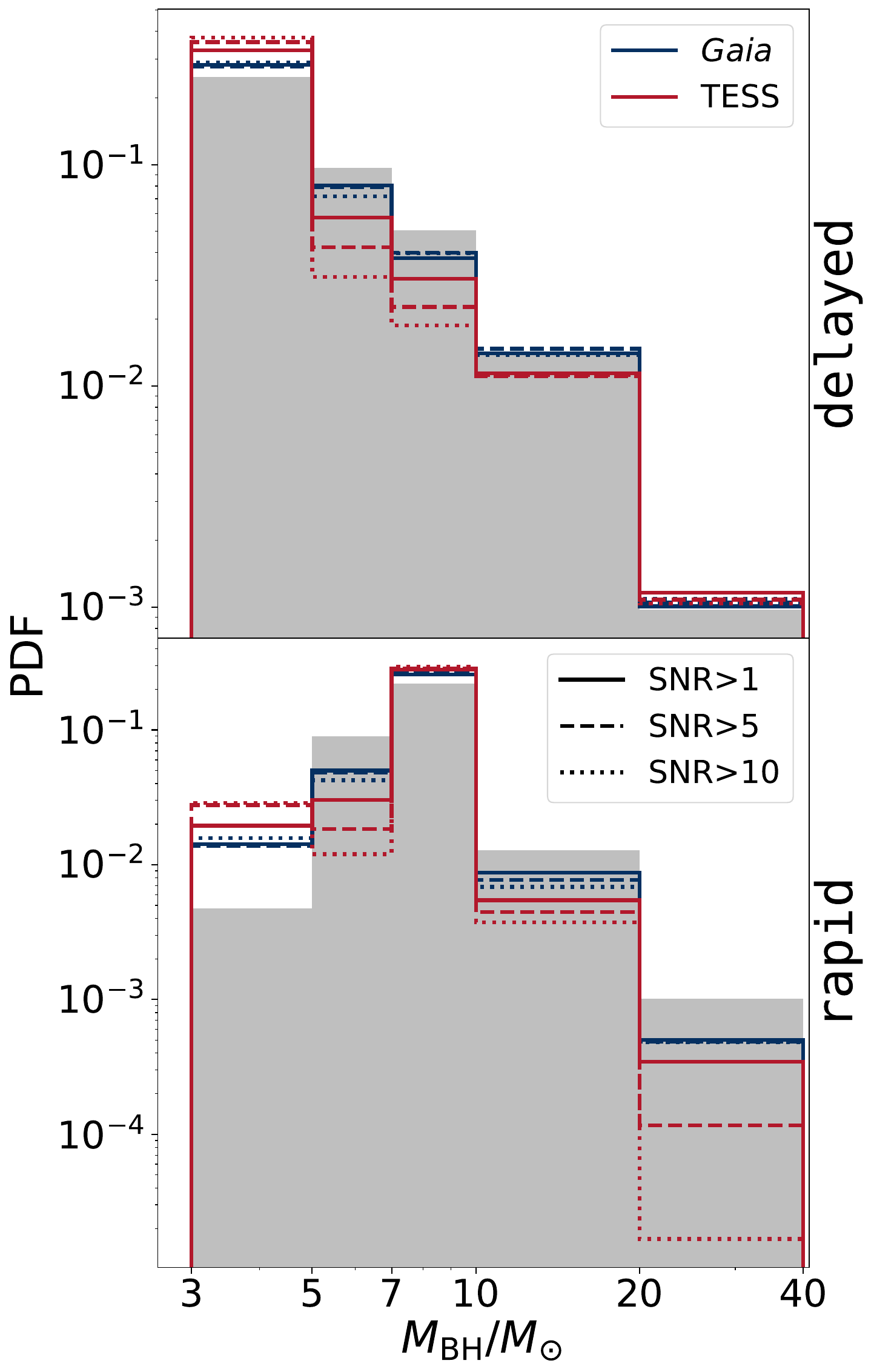}
    \caption{Same as the top panel of \autoref{fig:detection-frac-vs-mbh-ev} but includes BH--LC binaries with SNR$\ge1$ (solid}), $5$ (dashed), and $10$ (dotted). 
    \label{fig:mbh-multi-snr}
\end{figure}

\begin{figure}[h!]
    \plotone{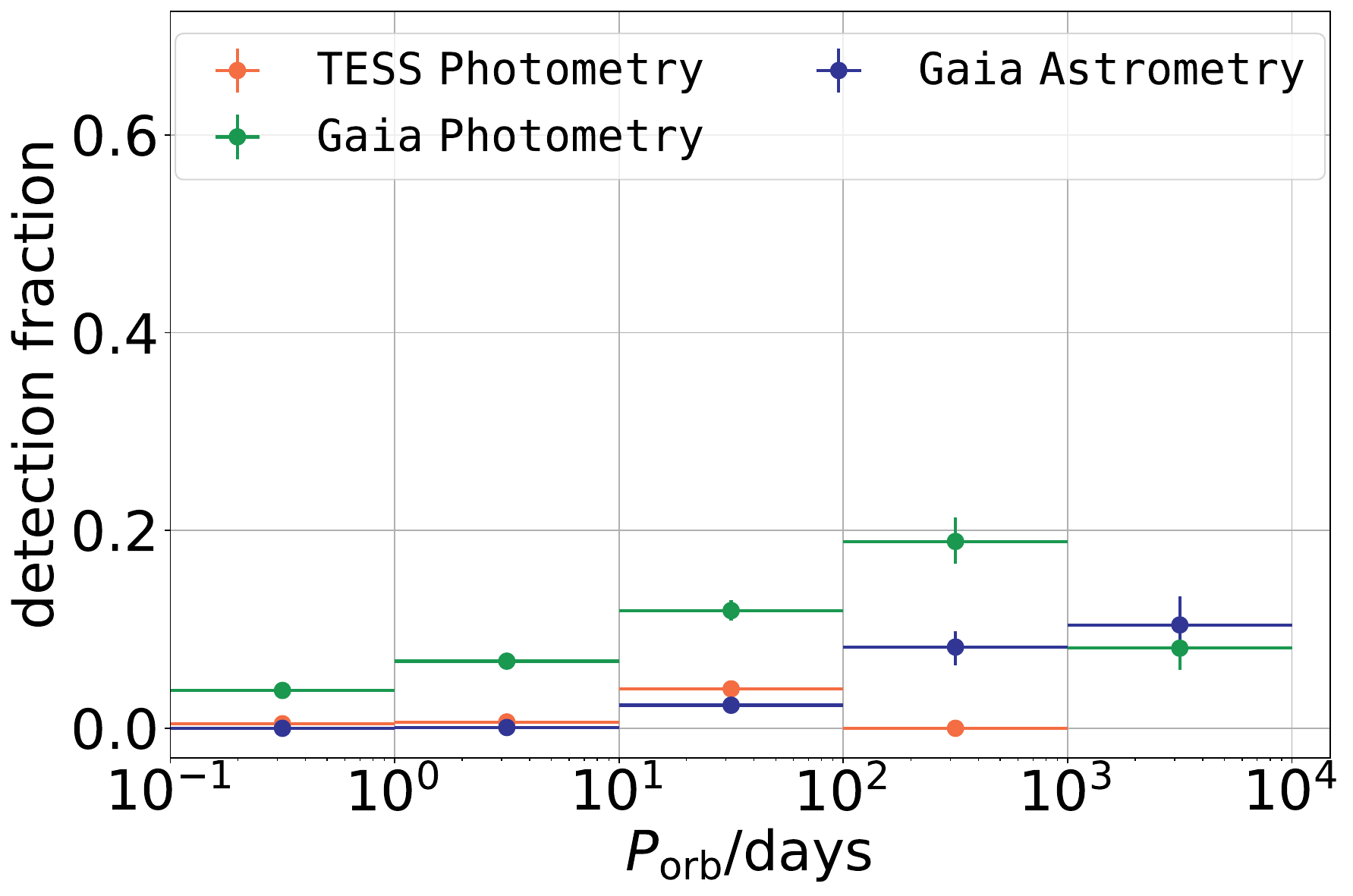}
    \caption{Same as \autoref{fig:detection-frac-vs-porb} but for the BH-LCs in our \delayed\ model. 
    }
    \label{fig:detection-frac-vs-porb-delayed}
\end{figure}

\begin{figure}[h!]
    \plotone{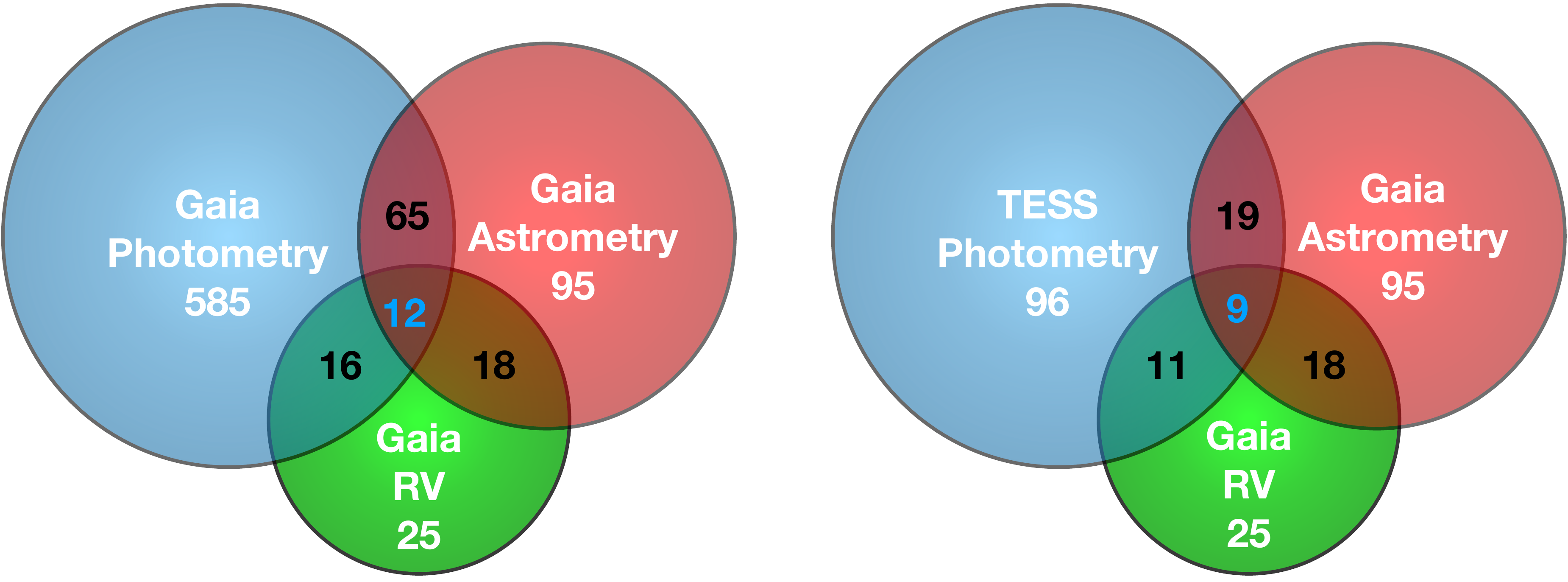}
    \caption{Same as \autoref{fig:gaia-rv-tess-obs} but for the BH-LCs in our \delayed\ model. 
    }
    \label{fig:gaia-rv-tess-obs-delayed}
\end{figure}

Here we present selected results from our \delayed\ model.
\autoref{fig:detection-frac-vs-porb-delayed} shows the detection fraction as a function of $\porb$ for the \delayed\ model for our detected population of BH-LC binaries. The detection fraction in the \delayed\ model is very similar to the same for the \rapid\ model (\autoref{fig:detection-frac-vs-porb}). \autoref{fig:ev_rb_sl_overlap-delayed} and \ref{fig:gaia-rv-tess-obs-delayed} shows the set diagram for the \delayed\ model. \autoref{fig:mbh-multi-snr} shows the detection fraction as a function of $\mbh$ for SNR $\ge$1, 5 and 10.

\end{document}